%% file: ms.tex
\documentclass[]{elsarticle}
\usepackage[dvipsnames]{xcolor}

\makeatletter
\def\ps@pprintTitle{%
 \let\@oddhead\@empty
 \let\@evenhead\
 \def\@oddfoot{}%
 \let\@evenfoot\@oddfoot}
\makeatother

\usepackage{amssymb}
\usepackage{amsmath}
\usepackage{textcomp}
\usepackage{mathtools}
\usepackage{subcaption}
\usepackage[section]{placeins}
\usepackage{comment}
\includecomment{comment}

\usepackage{lineno}

\usepackage{booktabs}

\usepackage[numbers]{natbib}
\usepackage{listings}

\input{packages}

\usepackage[dvipsnames]{xcolor}
\colorlet{Reviewer1}{BlueViolet}
\colorlet{Reviewer2}{ForestGreen}

\usepackage{xspace}
\usepackage{textcomp}
\newcommand{\unitspace}{\,}
\newcommand{\negativeunitspace}{\!}
\newcommand{\meter}{\unitspace{}m\xspace}
\newcommand{\squaremeter}{\unitspace{}m\textsuperscript{2}\xspace}
\newcommand{\cubicmeter}{\unitspace{}m\textsuperscript{3}\xspace}
\newcommand{\second}{\unitspace{}s\xspace}

\newcommand{\meterpersecond}{\unitspace{}m/s\xspace}
\newcommand{\squaremeterpersecond}{\unitspace{}m\textsuperscript{2}/s\xspace}
\newcommand{\newtonpermeter}{\unitspace{}N/m\xspace}
\newcommand{\pascalsecond}{\unitspace{}Pa\,s\xspace}
\newcommand{\kilogramspercubicmeter}{\unitspace{}kg/m\textsuperscript{3}\xspace}
\newcommand{\degreec}{\unitspace{}\textdegree{}C\xspace}

\newcommand{\expo}[2]{#1\cdot10^{#2}\xspace}

\allowdisplaybreaks

\begin{document}

\frenchspacing




\begin{frontmatter}



\author[1]{Anja Lippert \corref{corr}}
\ead{Anja.Lippert@bosch.com}
\cortext[corr]{Corresponding author. }
\author[1]{Tobias Tolle}
\ead{Tobias.Tolle@bosch.com}
\author[1]{Aaron D\"orr}
\ead{Aaron.Doerr@bosch.com}
\address[1]{Robert Bosch GmbH, Corperate Research, Robert-Bosch-Campus 1, 71272 Renningen, Germany}
\author[2]{Tomislav Maric}
\ead{maric@mma.tu-darmstadt.de }
\address[2]{Mathematical Modeling and Analysis, Technische Universität Darmstadt, Alarich-Weiss-Str. 10, 64287 Darmstadt, Germany}

\title{A benchmark for surface-tension-driven incompressible two-phase flows}


\input{sections/abstract.tex}


\end{frontmatter}


\input{sections/introduction.tex}

\input{sections/numerical-methods.tex}
\input{sections/results_convection}
\input{sections/results_hydrodyn}

\input{sections/conclusions}


\bibliographystyle{elsarticle-num-names} 
\bibliography{literature}


\end{document}

%% file: packages.tex
\usepackage{hyperref}
\usepackage{floatrow}
\usepackage{algorithm}
\usepackage{graphicx}
\usepackage{algpseudocode}
\usepackage{amsmath}
\usepackage[size=footnotesize]{subcaption}
\usepackage[textsize=tiny,disable]{todonotes}
\graphicspath{{figures/}}
\usepackage{cleveref}
\crefname{appendix}{}{}
\usepackage{soul}
\usepackage{geometry}
\usepackage{algorithmicx}
\usepackage{booktabs}
\usepackage{longtable}
\usepackage{siunitx}
\usepackage{stmaryrd}
\usepackage{algpseudocode}
\usepackage{algorithm}
\usepackage{adjustbox}
\usepackage{color, colortbl}
\usepackage{multirow}
\usepackage{xcolor}
\usepackage{ulem}
\usepackage{pgfgantt}
\usepackage{microtype}

\renewcommand{\v}{\mathbf{v}}
\newcommand{\x}{\mathbf{x}}
\newcommand{\n}{\mathbf{n}}

%% file: sections/abstract.tex
\begin{abstract}
The Volume-of-Fluid (VoF) method for simulating incompressible two-phase flows is widespread in academic and commercial simulation software because of its many advantages: a high degree of volume conservation, applicability to unstructured domain discretization (relevant for engineering applications), straightforward parallel implementation with the domain-decomposition and message-passing approach (important for large-scale simulations), and intrinsic handling of strong deformations and topological changes of the fluid interface. However, stable and accurate handling of small-scale capillary flows (dominated by surface tension forces) is still challenging for VoF methods. With many different VoF methods making their way into commercial and open-source software, it becomes increasingly important to compare them quantitatively and directly. For this purpose, we propose a set of simulation benchmarks and use them to directly compare VoF methods available in OpenFOAM, Basilisk, and Ansys Fluent. We use Jupyter notebooks to document and process the benchmark results, making a direct comparison of our results with other two-phase simulation methods very straightforward. The publicly available input data, secondary benchmark data, and post-processing Jupyter notebooks can be re-used by any two-phase flow simulation method that discretizes two-phase Navier-Stokes equations in a one-fluid formulation, which can save a significant amount of person-hours.
\end{abstract}

%% file: sections/introduction.tex
\section{Introduction}
\label{sec:intro}

Immiscible multiphase flows on small scales become increasingly important not only in academic research but also in industrial applications, e.g. in lab-on-a-chip applications or for ensuring media tightness for corrosion prevention in electrical components. Industrial multiphase flow processes generally occur in geometrically complex flow domains making predictive and accurate numerical methods in such domains essential. The Volume-of-Fluid (VoF) method is widely used because it allows for geometrically complex flow domains, large deformations and breakup/merging of the fluid interface, while ensuring a high degree of volume conservation. 

The VoF method dates back to \citeyear{hirt1981volume} \citep{hirt1981volume} and since then a multitude of different variants of the VoF method have emerged, distinguishable by different methods used to approximate the fluid interface and advect it. Geometrical VoF methods, although originally found to be computationally expensive, are emerging nowadays in both efficient and accurate versions and finding their use in commercial and open-source Computational Fluid Dynamics (CFD) software. Since the geometrical VoF methods are substantially more accurate than algebraic methods, they are the focus of this manuscript.
%
For the un-split, geometric transport on unstructured meshes alone numerous different methods exist (cf. \cite{maric2020unstructured} for a recent review), showing very active ongoing research in this field. 
Another crucial and actively researched aspect of the VoF method is a force-balanced, accurate, and numerically stable discretization of the surface tension force, reviewed recently in \citep{popinet2018numerical}. 

Comparing VoF methods directly and efficiently becomes increasingly relevant, with many different VoF method implementations becoming available in open-source and commercial software. Direct method comparison based on verification and validation tests \citep{Oberkampf2010verification} is also relevant for developing a specific VoF method - in that case, a comparison between an improved version and an existing version is necessary. To facilitate VoF method comparison for small-scale engineering problems, we adapt canonical verification and validation cases to include fluid combinations and dimensions relevant to industrial applications. This study captures the status quo of state-of-the-art VoF method implementations and sheds light on open challenges. The automated post-processing that relies on open-source software and a specification of an open secondary-data format provides a basis for continuous benchmarking \citep{Grambow2019} for two-phase flow simulation methods. 

Six different VoF method implementations are benchmarked for advection and hydrodynamic accuracy, namely: the interFoam and interIsoFoam OpenFOAM \citep{OpenFOAMcode} solvers, the interFlow solver from the OpenFoam sub-project TwoPhaseFlow \cite{scheufler2021twophaseflow,TwoPhaseFlowCode}, as well as the Ansys Fluent \citep{ansysfluentuser2020r1} and Basilisk~\citep{basilisk} software. All test cases are resolved with three increasing mesh refinements to test for mesh convergence, and the hydrodynamic cases are simulated with three different realistic fluid/fluid pairings.

To enable future comparisons of these and other methods, the benchmarks (results, setups and postprocessing  scripts) are made publicly available. The snapshot of the benchmark repository used for this publication is published at the TUDatalib data repository at TUDarmstadt \citep{BenchmarkData}, and the research repository is available at Bosch Research GitHub \citep{BoschGitHub}.

%% file: sections/numerical-methods.tex
\section{The Volume-of Fluid Method}
\label{sec:numerical-methods}

\subsection{Two-phase Navier-Stokes equations in the one-field 
formulation }

\begin{figure}[!htb]
    \centering
    \def\svgwidth{0.6\textwidth}
    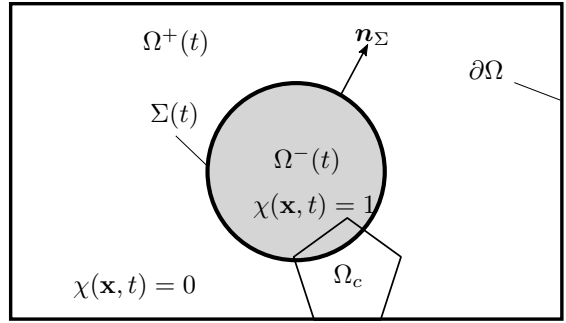\notag
    \caption{Two phase flows domain: the ellipsoidal
 heavy phase flow is surrounded by the light phase flow.}
    \label{fig:drop-gas}
\end{figure}

\label{sec:volume-of-fluid}
The Volume of Fluid (VoF) method utilizes a one-field formulation of the Navier-Stokes Equations (NSE) for two-phase flows. 

The solution domain (cf. \cref{fig:drop-gas}) $\Omega \subseteq \mathbb{R}^d, d = 1,2,3$, is separated into two evolving subdomains $\Omega^\pm(t) \subseteq \Omega$ each occupied with a fluid phase "$+$" or "$-$" with respective constant densities $\rho^\pm$ and viscosities $\mu^\pm$, separated by the fluid interface $\Sigma(t):=\partial \Omega^-(t), \Sigma(t) \in \mathbb{R}^{d-1}$.  
The phase indicator
\begin{align}
\chi\left(\mathbf{x}, t\right)=
    \begin{cases}
        1 & \text{if } x \in \Omega^-(t), \\
        0 & \text{if } x \in \Omega^+(t)  
    \end{cases}
    \label{eq:phaseind}
\end{align}
indicates fluid phases $\Omega^{\pm}(t)$. 
The phase-indicator defines the single-field density and viscosity
\begin{equation}
 \begin{aligned}
    \rho(\x, t)  = \chi(\x, t) \rho^- + (1 - \chi(\x, t)) \rho^+, \\
    \mu(\x, t)  = \chi(\x, t) \mu^- + (1 - \chi(\x, t)) \mu^+,
 \end{aligned}
 \label{eq:dens-visc}
\end{equation}
with $\rho^\pm,\mu^\pm$ as constant densities and viscosities of fluid phases occupying $\Omega^\pm$, respectively. 

Assuming the incompressibility of both fluid phases results in 
\begin{equation}
    \nabla\cdot \v = 0
    \label{eq:volconserv}
\end{equation}
and \cref{eq:dens-visc}, together with the surface-tension force responsible for the momentum transfer at $\Sigma(t)$, results in the single-field formulation of the two-phase momentum conservation equation 
\begin{equation}
    \partial_t(\rho \v) + \nabla\cdot(\rho \v \v) - \nabla\cdot(\mu \nabla \v) = -\nabla p + \rho\mathbf{g} + \mathbf{f}_\Sigma + \nabla\cdot(\mu (\nabla \v)^T), 
    \label{eq:momentum}
\end{equation}
where 
\begin{equation}
    \mathbf{f}_\Sigma:=\sigma \kappa_\sigma \n_\Sigma \delta_\Sigma
    \label{eq:fsigma}
\end{equation}
is the surface tension force with the surface tension coefficient $\sigma$, mean curvature of the fluid interface $\Sigma(t)$ $\kappa_\Sigma$, $\n_\Sigma$ the interface-normal (cf. \cref{fig:drop-gas}) and the interface Dirac's distribution $\delta_\Sigma$ that activates $\mathbf{f}_\Sigma$ only at $\Sigma(t)$. More information on the mathematical model is available in \citep{Tryggvason2011}.

For arbitrary initial and boundary conditions the exact solution to one-field formulation of two-phase Navier-Stokes equations given by \cref{eq:volconserv,eq:momentum} cannot be obtained. 
The sharpness of the phase indicator \ref{eq:phaseind} further complicates numerical solutions of \cref{eq:volconserv,eq:momentum}  by algebraic discretization methods.

Equivalent to the unstructured Finite Volume Method \citep{Jasak1996,Hirsch2007,Maric2014,Moukalled2016}, the VoF method decomposes $\Omega$ into a set of non-overlapping fixed-in-time control volumes $\Omega_c$
\begin{equation}
    \Omega = \bigcup_{c \in C} \Omega_c, \quad \Omega_c \cap \Omega_k \in \mathbb{R}^e, e<d, \forall c,k \in C, c \ne k.
\end{equation}
For a fixed-in-time control volume $\Omega_c$, the volume fraction $\alpha_c(t)$ is defined as
\begin{align}
    \alpha_c(t) :=\frac{1}{|\Omega_c|}\int_{\Omega_c} \chi(\mathbf{x}, t) \, dV\,
    \label{eq:alphadef}
\end{align}
and it represents the fractional volume of the phase $\Omega^-(t)$ occupying $\Omega_c$ at time $t$. 
Incompressible flow condition~\labelcref{eq:volconserv} is equivalent to volume conservation in each fluid phase $\Omega^{\pm}$. Expressing the rate of change of any of the phase-specific volumes
\begin{equation}
\begin{aligned}
    \Omega_c^-(t) := \Omega^-(t) \cap \Omega_c = \int_{\Omega_c} \chi(\x, t) \, dV,\\
   \Omega_c^+(t) := \Omega^+(t) \cap \Omega_c = \int_{\Omega_c} (1 - \chi(\x, t)) \, dV, \\
\end{aligned}
\end{equation} 
say, $\Omega_c^-(t)$, using the Reynolds transport theorem (RTT), and dividing by $|\Omega_c|$ leads to
\begin{equation}
    \partial_t \alpha_c(t) = -\frac{1}{|\Omega_c|}\int_{\partial \Omega_c} \chi(\x, t) \v \cdot \n \, dS,
    \label{eq:alphartt}
\end{equation}
an exact equation if 
\begin{equation}
    \bigcup _{c \in C} (\partial \Omega_c \cap \partial \Omega) = \partial \Omega,
    \label{eq:exactboundary}
\end{equation}
which is frequently violated, e.g., if $\partial \Omega$ is non-linear and it is discretized by boundaries of control-volumes that are linear.

\subsection{Discretization}
\label{sec:discretization}

Since the goal of establishing a benchmark is to establish expectations in terms of accuracy on different methods irrespective of their differences, we only briefly cover the differences between VOF methods used here in direct comparison. The reader is pointed to other sources of literature for further details.

\subsubsection{Volume fraction advection}

The VoF method was developed into two different categories over the years, the algebraic and geometric VoF methods, both start with integrating \cref{eq:alphartt} over a time step $[t^n, t^{n+1}]$, and a fixed polyhedral control volume, whose boundary is bounded by possibly non-convex and non-planar faces $\partial \Omega_c =\bigcup_{f \in F_c} S_f$, leading  to
\begin{equation}
    \alpha_c^{n+1} = \alpha_c^n - \frac{1}{|\Omega_c|}\sum_{f \in F_c} \int_{t^n}^{t^{n+1}} \int_{S_f} \chi(\x, t) \v \cdot \n \, dS \, dt, 
    \label{eq:alphaadvect}
\end{equation}
using the notation $\phi_c^{n} \equiv \phi_c(t=t^n)$. The volume fraction advection equation \ref{eq:alphaadvect} is still exact.

\paragraph{Algebraic VoF methods}

Unstructured algebraic VOF methods use
\begin{equation}
    \alpha_f := \frac{1}{|S_f|}\int_{S_f} \chi(\x,t) \, dS
    \label{eq:alphaf}
\end{equation}
in \cref{eq:alphaadvect}, and face-centered FV velocity approximation $\mathbf{v}_f$ to obtain
\begin{equation}
\begin{aligned}
    \alpha_c^{n+1} & = \alpha_c^n - \frac{1}{|\Omega_c|}\sum_{f \in F_c} \int_{t^n}^{t^{n+1}} \mathbf{v}_f(t) \cdot \frac{\mathbf{S}_f}{|S_f|}  \int_{S_f} \chi(\x, t) \, dS \, dt, \\
    & = \alpha_c^n - \frac{1}{|\Omega_c|}\sum_{f \in F_c} \int_{t^n}^{t^{n+1}} F_f(t) \alpha_f(t) dt. \\
    \label{eq:alphaalgebr}
\end{aligned}
\end{equation}
The temporal integration quadrature used the phase-specific volumetric flux determines the temporal accuracy. For example, the trapezoidal quadrature results in a Crank-Nicolson discretization of \cref{eq:alphartt}, 
\begin{equation}
     \alpha_c^{n+1} - \alpha_c^n + 0.5 \frac{\Delta t}{|\Omega_c|} (F_f^{n+1}\alpha_f^{n+1} + F_f^{n} \alpha_f^n) = 0.
     \label{eq:alphacrank}
\end{equation}
All algebraic methods rely on algebraic approximations of $\alpha_f=\alpha_f(\alpha_{Of}, \alpha_{Nf})$, which is less accurate than the geometric approach. Still, algebraic approximation of $\alpha_f$ is computationally much faster than the geometric integration of \cref{eq:alphartt} discussed below. Furthermore, algebraic methods allow for an implicit discretization, shown e.g. in \cref{eq:alphacrank}, leading to increased numerical stability \citep{Denner2022implicit}.
Algebraic methods suffer from dispersion errors. The phase-specific volumetric fluxes $F_f \alpha_f$ must be limited to suppress these errors \citep{Deshphande2012}. Some methods extend \cref{eq:alphartt} with an anti-diffusion (compression) term in the attempt to maintain sharpness of $\{\alpha_c\}_{c \in C}$ \citep{Deshphande2012}, others recover a hyperbolic tangent in the interface-normal direction from $\{\alpha_c\}_c$ \citep{Chen2022}. 

\paragraph{Geometrical VoF methods}

Geometrical VoF methods are called geometrical, because they discretize \cref{eq:alphartt} by geometrically approximating $\chi(\x,t)$ on the r.h.s. of \cref{eq:alphartt}, and integrate \cref{eq:alphartt} in time using geometrical calculations. 
Unstructured geometrical VoF methods also compute the r.h.s integral of \cref{eq:alphartt} over a finite volume whose boundary $\partial_c$ consists of possibly non-planar faces (surfaces), that are, however, bounded with edges (mesh edges) $\partial_c \Omega_c := \bigcup_{f \in F_c} S_f$. 

We avoid repeating the details of approximate solutions of \cref{eq:alphaadvect} by different VoF methods used in this benchmark, the reader is directed to \citep{maric2020unstructured} for a recent review of unstructured geometrical VoF methods, \citep{Roenby2016,Scheufler2019,scheufler2021twophaseflow} for the OpenFOAM solvers implementing plicRDF-isoAdvector schemes, \citep{popinet2009} for Basilisk, and \citep{ansysfluenttheory2020r1} for Ansys Fluent.

It suffices to know that every geometrical VoF method integrates the r.h.s. of \cref{eq:alphaadvect} differently, but all methods result with a discrete equation of the form
\begin{equation}
     \alpha_c^{n+1} = \alpha_c^n - \frac{1}{|\Omega_c|}\sum_{f \in F_c} V_f^\alpha, 
     \label{eq:alphavfalpha}
\end{equation}
where $V_f^\alpha$ is the phase-specific volume fluxed through the face $S_f\in \partial \Omega_c$ over the time step $[t^n, t^{n+1}]$.

Volume conservation, monotonicity, stability, boundedness-preserving property ($\alpha_c \in [0,1]$ by definition~\labelcref{eq:alphadef}), serial and parallel computational efficiency (numerical consistency in parallel), absolute accuracy and convergence order, ability to handle geometrically complex solution domains~$\Omega$, are all determined by the way a VoF method approximates $V_f^\alpha$ \citep{maric2020unstructured}. Methods benchmarked here all approximate $\chi(\x,t)$ using a piecewise-linear approximation of~$\Sigma_c(t) := \Sigma \cap \Omega_c$, the part of the fluid interface in a cell $\Omega_c$. This Piecewise Linear Interface Calculation, PLIC, (cf. \citep{maric2020unstructured} for details) approximates $\Sigma_c(t)$ as a plane in each cell $\Omega_c$ intersected by the fluid interface $\Sigma(t)$.  

Errors in $V_f^\alpha$ discretization, even if they are very small, may lead to very large errors in $\alpha_c^{n+1}$. This becomes evident from \cref{eq:alphavfalpha} because $|\Omega_c| \propto h^d$ ($d$ is the spatial dimension) divides any residual geometrical error in $V_f^\alpha$. This error amplification can lead to catastrophic failure in the approximation of the surface tension force $\mathbf{f}_\sigma$ and is still challenging in some cases for structured geometrical VoF methods, and even more so for unstructured VoF methods.

\subsubsection{Surface tension force}

All VoF methods benchmarked in \cref{sec:benchmark} model the surface tension force using \cref{eq:fsigma}. 
For small-scale problems, surface tension force plays a crucial role, making its correct discretization necessary to achieve accurate simulation results. The surface tension force is discretized as a volumetric force density in the vicinity of the interface using the Continuum Surface Force (CSF) model \cite{brackbill1992continuum}
\begin{align}
\mathbf{f}_{\Sigma,f} = \sigma\kappa_{\Sigma_f}(\nabla\alpha)_f,
\end{align}
averaged over $S_f$ at the face-centroid, where $(\nabla\alpha)_f$ denotes the discrete gradient operator, and $(\nabla \alpha)_f \approx \n_\Sigma \delta_\Sigma$ in \cref{eq:fsigma}.

The main difference between different VoF methods lies in the approximation of the mean curvature~$\kappa$. The simplest VoF curvature model is
\begin{align}
    \kappa_{\Sigma, f} =-\left[\nabla\cdot\left(\frac{\nabla \alpha}{|\nabla \alpha|}\right)\right]_f\,.
    \label{eq:csf-curvature}
\end{align}
which is obviously problematic to evaluate on finer meshes, since $|\nabla \alpha| \to \delta_\Sigma \n_\Sigma$ when $h \to 0$, $h$ being the discretization length.
For structured, hexahedral meshes, \citet{torrey1985} proposed a height-function based curvature approximation
\begin{align}
    \kappa= \frac{h''}{\left(1+h'^2\right)^{3/2}}\quad\text{(2D formulation)},
    \label{eq:hf-curvature}
\end{align}
which is used in Basilisk \cite{basilisk} and its
predecessor Gerris \cite{popinet2009}.
While well-suited for Cartesian meshes, this approach poses considerable challenges for unstructured, polyhedral meshes, in terms of complexity and computational costs \cite{ivey2015}.

Another approach for curvature approximation is to locally fit a paraboloid to an interface cell and its neighbour cells. Curvature is then analytically calculated using the paraboloid parameterization. Fitting is formulated as a least squares problem to best match e.g. volume fractions \cite{jibben2019} or
centroids of PLIC polygons \citep{scheufler2021twophaseflow}.

The last method used by the plicRDF-isoAdvector method \citep{Scheufler2019,scheufler2021twophaseflow} is to compute the mean
curvature from a Reconstructed Distance Function, RDF, \cite{cummins2005}. In this approach, a signed distance $\phi$ is reconstructed geometrically from the PLIC elements.
From $\phi$, the curvature is calculated analogously to \cref{eq:csf-curvature} by
\begin{equation}
    \kappa = -\nabla \cdot \left( \frac{\nabla\phi}{|\nabla\phi|} \right).
    \label{eq:rdf-curvature}
\end{equation}

\subsection{Benchmarked methods}
\label{sec:solver-overview}
The six different solvers listed in \cref{tab:solver-overview} are considered for the test cases in \cref{sec:benchmark}. \Cref{tab:solver-overview} shows their main attributes, ranging from licensing to discretization of phase advection and surface tension approximation.
The solvers are chosen based on their common usage in either industry or academia or both.
For all solvers we use the recommended settings and do not perform any use-case-specific adaptation which might taint the results. 

\begin{table}[ht]
\footnotesize
    \begin{tabular}{ p{2cm}p{2cm}p{2cm}p{2cm}p{2cm}>{\raggedright\arraybackslash}p{2cm}}
     \toprule
     Solver &  Publisher & license & mesh & $\alpha$-transport & curvature appr. \\
    \midrule
     interFoam & OpenCFD Ltd & Gnu GPL v3 & unstructured & algebraic VoF & derivative (\cref{eq:csf-curvature})\\
     interIsoFoam & OpenCFD Ltd & Gnu GPL v3 & unstructured & geom. VoF & derivative (\cref{eq:csf-curvature}) \\
     TwoPhaseFlow & - & Gnu GPL v3 & unstructured & geom. VoF  & parabolic fit\\
    TwoPhaseFlow & - & Gnu GPL v3 & unstructured & geom. VoF  & reconstr. distance funct. (\cref{eq:rdf-curvature}) \\
     Fluent & Ansys Inc. & commercial & unstructured & geom. VoF &  derivative \\
     Basilisk & - & Gnu GPL v3  & structured & geom. VoF  & height funct. (\cref{eq:hf-curvature})\\
    \bottomrule
    \end{tabular}
    \caption{Overview of benchmarked solvers and some relevant properties.}
    \label{tab:solver-overview}
\end{table}
All present solvers offer the additional capability of adaptive mesh refinement, however, it was not applied in this study.
In addition, except for Basilisk, all tools allow polyhedral meshes. While beneficial for complex geometries as present in industrial applications, this additional layer of numerical complexity is not considered here. Hence, only Cartesian meshes are used in the following \cref{sec:benchmark}. Furthermore, regarding mesh quality Cartesian meshes are optimal for
the Finite Volume discretization used by the benchmarked solvers \cite{juretic2005phd}.
For OpenFOAM-related cases OpenFOAM v2112 \cite{OpenFOAMcode} is used with
\verb+blockMesh+ for grid generation. The OpenFoam sub-project TwoPhaseFlow also uses \verb+blockMesh+.
Fluent was used with Ansys Meshing~\citep{ansysmeshinguser2020r1} to generate meshes while the geometries are constructed and exported as STEP file from either Gmsh~\cite{gmsh2020} in 2D or FreeCAD~\cite{freecad2022} in 3D.
Basilisk cases are build up with the Basilisk library \cite{basilisk}, the successor of Gerris \cite{popinet2009}.

%% file: figures/model_HDR.pdf_tex
\begingroup%
  \makeatletter%
  \providecommand\color[2][]{%
    \errmessage{(Inkscape) Color is used for the text in Inkscape, but the package 'color.sty' is not loaded}%
    \renewcommand\color[2][]{}%
  }%
  \providecommand\transparent[1]{%
    \errmessage{(Inkscape) Transparency is used (non-zero) for the text in Inkscape, but the package 'transparent.sty' is not loaded}%
    \renewcommand\transparent[1]{}%
  }%
  \providecommand\rotatebox[2]{#2}%
  \newcommand*\fsize{\dimexpr\f@size pt\relax}%
  \newcommand*\lineheight[1]{\fontsize{\fsize}{#1\fsize}\selectfont}%
  \ifx\svgwidth\undefined%
    \setlength{\unitlength}{586.24039111bp}%
    \ifx\svgscale\undefined%
      \relax%
    \else%
      \setlength{\unitlength}{\unitlength * \real{\svgscale}}%
    \fi%
  \else%
    \setlength{\unitlength}{\svgwidth}%
  \fi%
  \global\let\svgwidth\undefined%
  \global\let\svgscale\undefined%
  \makeatother%
  \begin{picture}(1,0.46784722)%
    \lineheight{1}%
    \setlength\tabcolsep{0pt}%
    \put(0,0){\includegraphics[width=\unitlength,page=1]{model_HDR.pdf}}%
    \put(0.38812574,0.22418563){\color[rgb]{0,0,0}\makebox(0,0)[lt]{\smash{\begin{tabular}[t]{l}$\Omega^-(t)$\end{tabular}}}}%
    \put(0.47580089,0.05610961){\color[rgb]{0,0,0}\makebox(0,0)[lt]{\smash{\begin{tabular}[t]{l}$\Omega_c$\end{tabular}}}}%
    \put(0.35794211,0.15849391){\color[rgb]{0,0,0}\makebox(0,0)[lt]{\smash{\begin{tabular}[t]{l}$\chi(\x, t) = 1$\end{tabular}}}}%
    \put(0.09414215,0.04154651){\color[rgb]{0,0,0}\makebox(0,0)[lt]{\smash{\begin{tabular}[t]{l}$\chi(\x, t) = 0$\end{tabular}}}}%
    \put(0.67440925,0.35546509){\color[rgb]{0,0,0}\makebox(0,0)[lt]{\smash{\begin{tabular}[t]{l}$\partial \Omega$\end{tabular}}}}%
    \put(0.50713779,0.41067165){\color[rgb]{0,0,0}\makebox(0,0)[lt]{\smash{\begin{tabular}[t]{l}$\boldsymbol{n}_{\Sigma}$\end{tabular}}}}%
    \put(0.19627072,0.39732333){\color[rgb]{0,0,0}\makebox(0,0)[lt]{\smash{\begin{tabular}[t]{l}$\Omega^+(t)$\end{tabular}}}}%
    \put(0.20682113,0.2912696){\color[rgb]{0,0,0}\makebox(0,0)[lt]{\smash{\begin{tabular}[t]{l}$\Sigma(t)$\end{tabular}}}}%
    \put(0,0){\includegraphics[width=\unitlength,page=2]{model_HDR.pdf}}%
  \end{picture}%
\endgroup%

%% file: sections/results_convection.tex
\section{Benchmark cases}
\label{sec:benchmark}
The following cases serve to quantify each VoF-implementation separately as well as compare the solvers described in Sec.\,\ref{sec:solver-overview}.

\subsection{Convection cases}
\label{sec:benchmark-convC}
The accuracy of the convective transport of the volume fraction field $\alpha$ is compared for three cases, namely 2D diagonal transport of a sphere, a sphere in a 2D vortex flow and a sphere in a 3D vortex flow.
All flows are reversed halfway though the simulation so that an easy comparison between the volume fraction field at the beginning and end is possible via subtraction.
The $L_1$ shape error is defined as
\begin{equation}
    e_\text{shape} = \sum_i^{N_\text{cells}}V_i\left|\alpha_i^\text{end} - \alpha_i^\text{start}\right|\,.
    \label{eq:shape-error}
\end{equation}
Note that for the two-dimensional test cases $V_i$ refers to cell areas and only in three dimensions to the cell volumes.
The time discretization of $\Delta t=\expo{1}{-5}$\second results in a CFL number below unity for all cases.
In addition, the two TwoPhaseFlow solvers are not tested separately for convection since they build on interIsoFoam. 

\subsubsection{2D sphere in diagonal flow}
To analyze grid-orientation dependence of transport, a disc of radius $R=0.25$\meter is initialized at $(x_0,y_0)=(0.5,0.5)$\meter and transported diagonally on a Cartesian mesh in a rectangular domain of 3\texttimes{}2\squaremeter with the background velocity $\mathbf{U}=\left(1,0.5\right)$\meterpersecond. The flow is reversed after 2\second. The domain is resolved with 90\texttimes{}60~cells and twice refined to capture the grid dependence.

\begin{figure}[htb]
    \centering
    \begin{subfigure}{0.5\textwidth}
        \includegraphics[width=\textwidth]{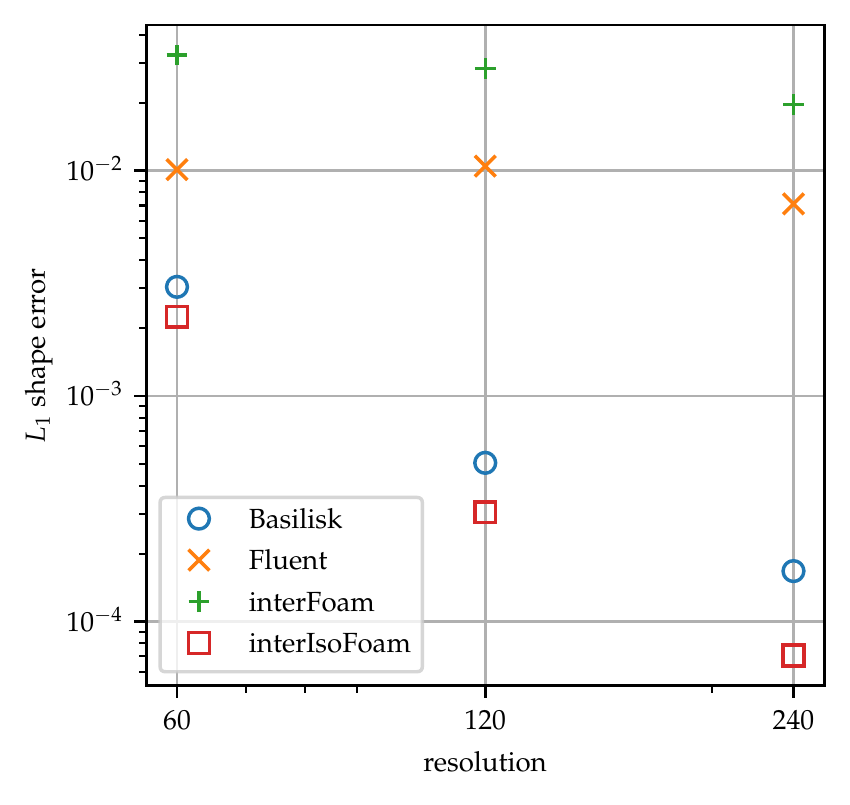}
        \subcaption{L1-shape error for 2D sphere in reversed diagonal flow and different resolutions}
        \label{fig:reversedDiagFlow2Da}
    \end{subfigure}
     \hspace{0.5cm}
    \begin{subfigure}{0.2\textwidth}
        \centering
        \includegraphics[width=0.8\textwidth]{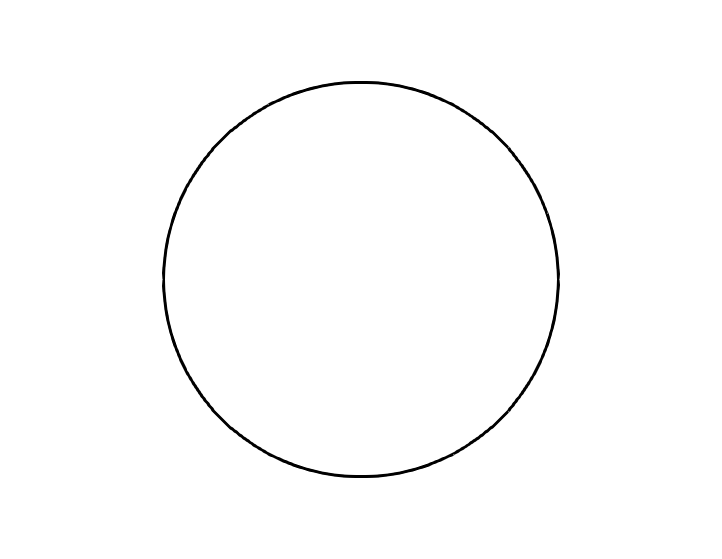}\\
        \includegraphics[width=0.8\textwidth]{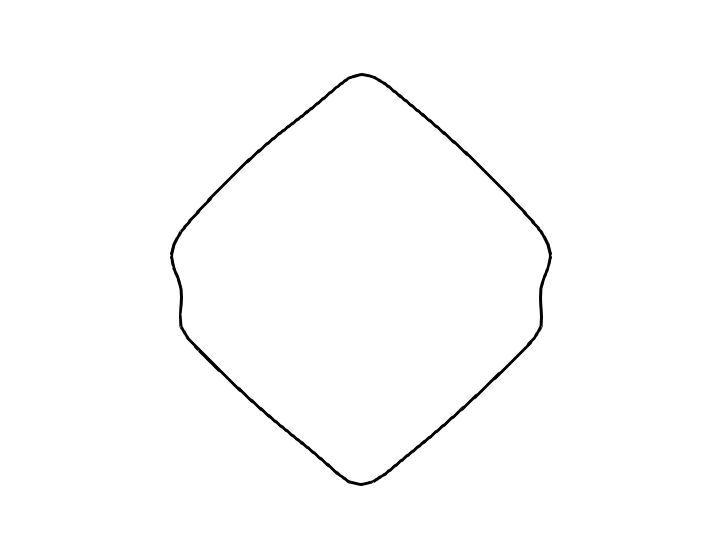}\\
        \includegraphics[width=0.8\textwidth]{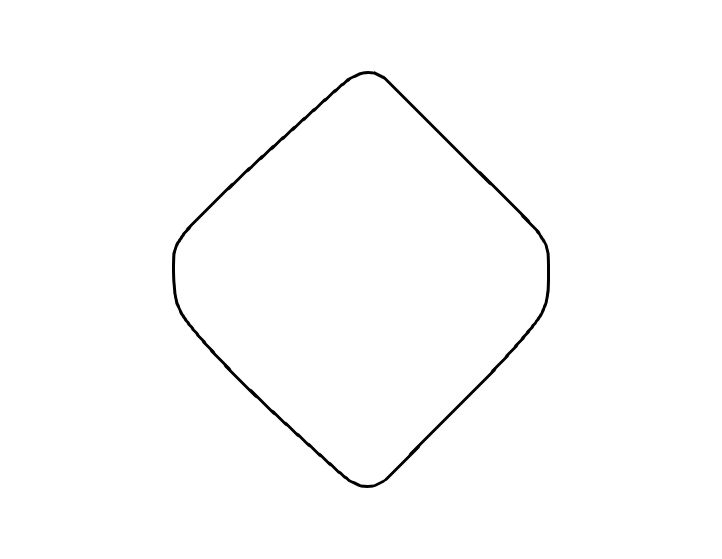}
        \subcaption{Isocontour of drop with interFoam for res.~120 at: top $t=0$\second, middle $t=2$\second, bottom $t=4$\second}
        \label{fig:reversedDiagFlow2Db}
    \end{subfigure}
    \hspace{0.5cm}
    \begin{subfigure}{0.2\textwidth}
        \centering
        \includegraphics[width=0.8\textwidth]{figures/convection_isosurfaces/discDiagonal2D/discInReversedDiagonalFlow_interFoam_interIsoFoam_n120_t0.png}\\
        \includegraphics[width=0.8\textwidth]{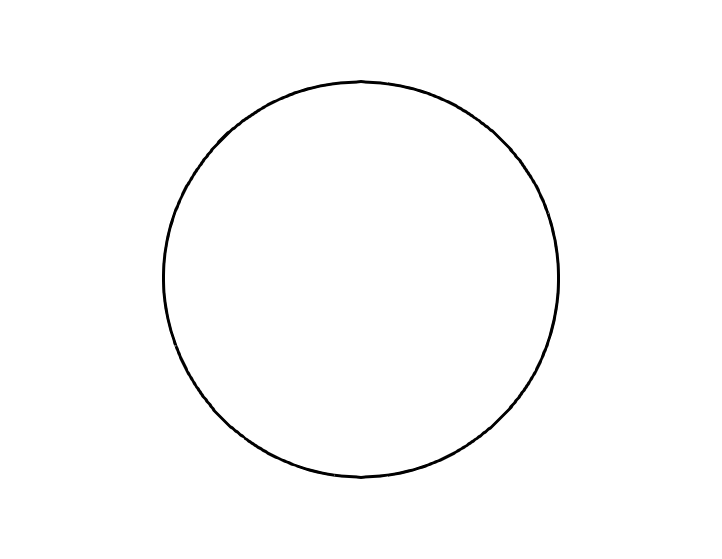}\\
        \includegraphics[width=0.8\textwidth]{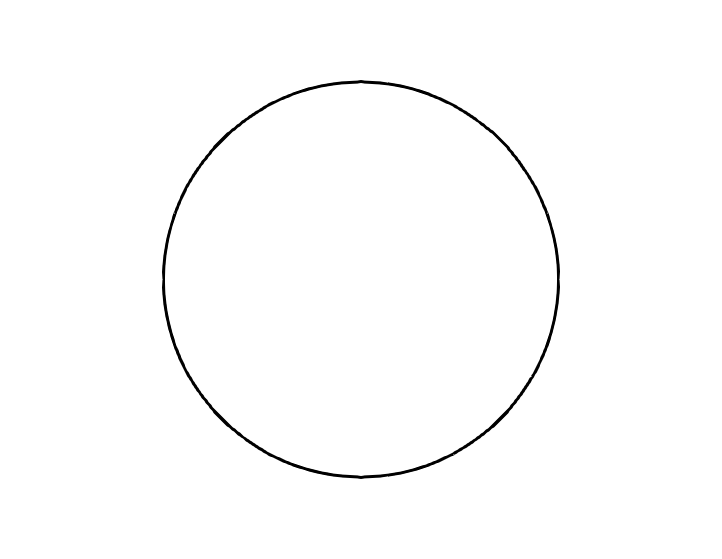}    
        \subcaption{Isocontour of drop with interIsoFoam for res.~120 at: top $t=0$\second, middle $t=2$\second, bottom $t=4$\second}
        \label{fig:reversedDiagFlow2Dc}
    \end{subfigure}   
    \caption{Error visualization for 2D sphere in reversed diagonal flow for different resolutions.}
\end{figure}
\Cref{fig:reversedDiagFlow2Da} shows the absolute shape error after 4\second of transport. For all codes the error overall reduces with increasing mesh resolution. interIsoFoam and Basilisk exhibit the smallest error as expected due to their geometrical transport. 
Surprisingly Fluent shows an higher error than expected, since a geometric transport is chosen here as well. The errors span with 1 to 2 orders of magnitude a much broader band than expected for such widely applied tools.
Visualizing the actual drop shape at the reversal point and at the end in \cref{fig:reversedDiagFlow2Db,fig:reversedDiagFlow2Dc} shows the actual "meaning" behind the shape error, showing a completely deformed interface contour. Determining any shape depending forces, for instance surface tension, will result in a large error.


\subsubsection{2D sphere in vortex flow}
To test the accuracy of convection against a complex flow in two-dimensions, a disc of radius $R=0.15$\meter is initialized at $(x_0,y_0)=(0.5,0.75)$\meter in a domain of 1\texttimes{}1\squaremeter. The velocity is prescribed~as
\begin{align}
u &=\cos\left( \pi t /t_{end}\right)\left(-2\sin^2(\pi x)\sin(\pi y)\cos(\pi y)\right)\\
v &=\cos\left( \pi t /t_{end}\right)\left(2\sin^2(\pi y)\sin(\pi x)\cos(\pi x)\right)
\end{align}
with an end time of $t_{end}=3$\second.
The domain is resolved by 32\texttimes{}32 cells and twice refined.
\begin{figure}[htb]
    \centering
    \begin{subfigure}{0.5\textwidth}
        \includegraphics[width=\textwidth]{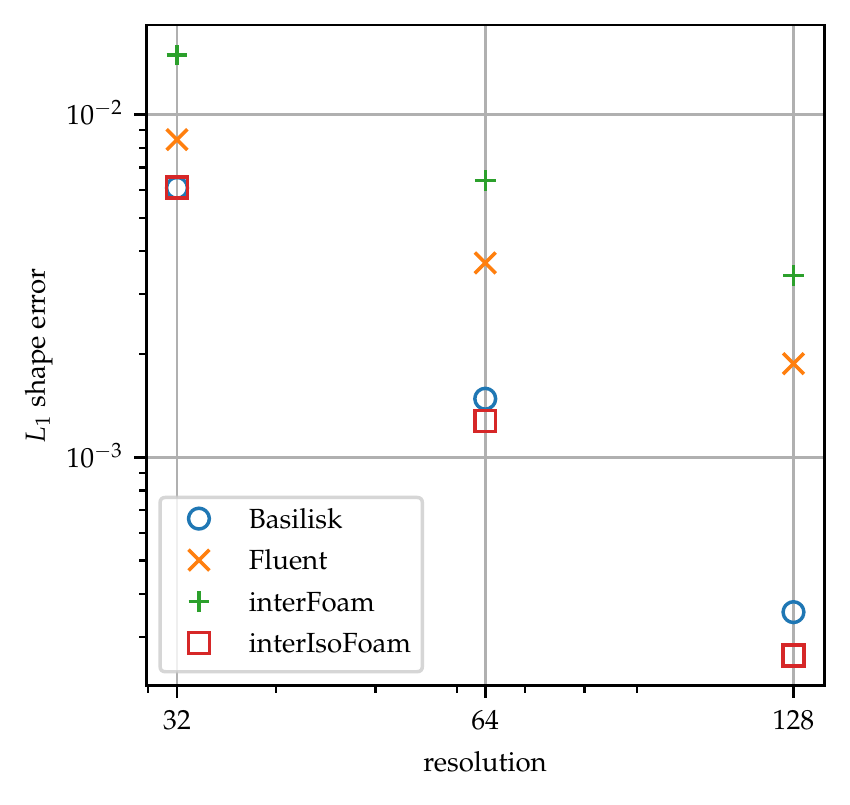}
        \subcaption{$L_1$ shape error for 2D sphere in reversed vortex flow and different resolutions.}
        \label{fig:reversedVortexFlow2Da}
    \end{subfigure}
    \hspace{0.5cm}
    \begin{subfigure}{0.2\textwidth}
        \centering
        \includegraphics[width=\textwidth]{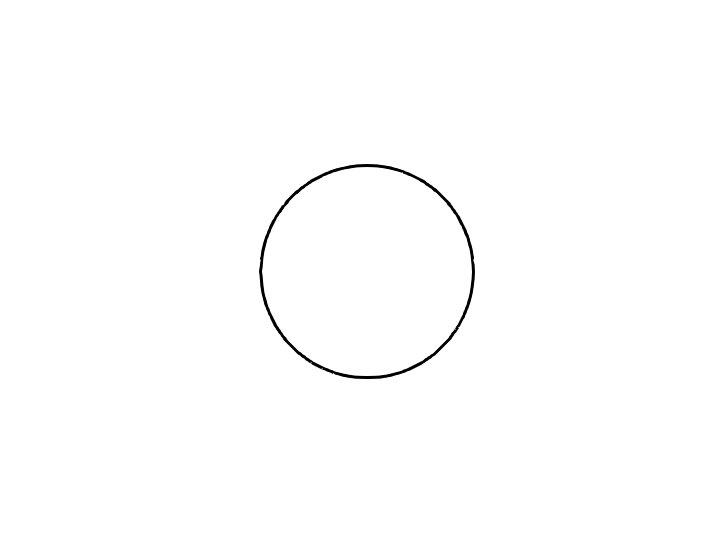}\\
        \includegraphics[width=\textwidth]{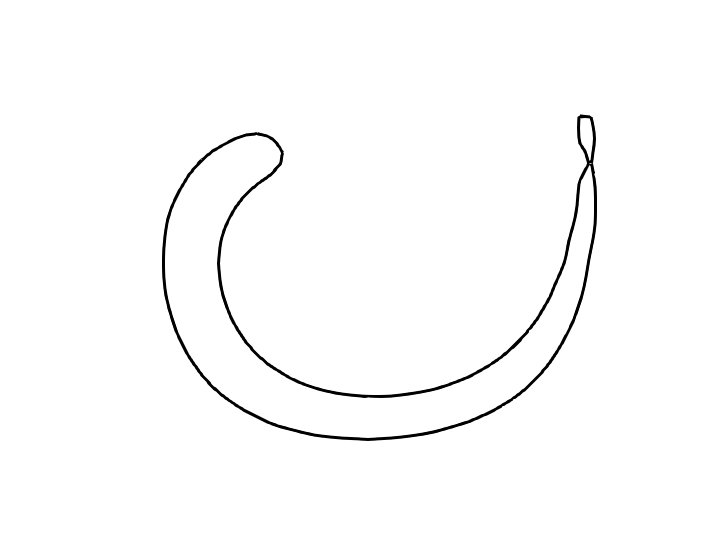}\\
        \includegraphics[width=\textwidth]{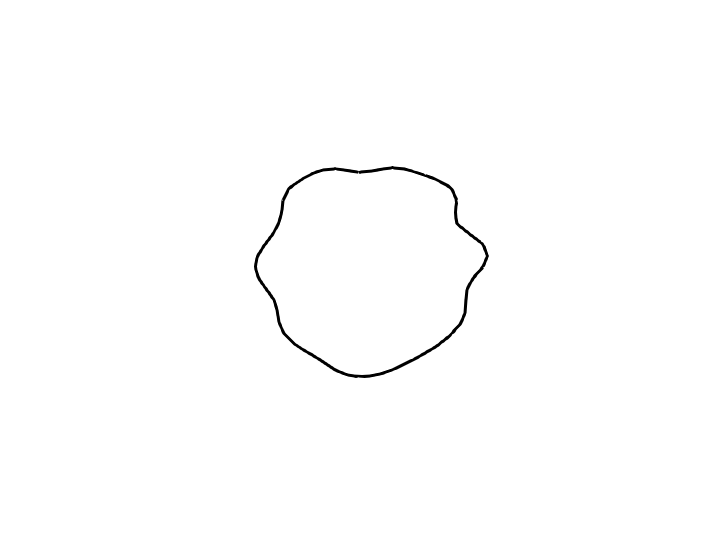}
        \subcaption{Isocontour of drop with interFoam for res.~64 at: top $t=0$\second, middle $t=1.5$\second, bottom $t=3$\second}
        \label{fig:reversedVortexFlow2Db}
    \end{subfigure}
    \hspace{0.5cm}
    \begin{subfigure}{0.2\textwidth}
        \centering
        \includegraphics[width=\textwidth]{figures/convection_isosurfaces/discVortex2D/discVortex2D_interFoam_interIsoFoam_n64_t0.png}\\
        \includegraphics[width=\textwidth]{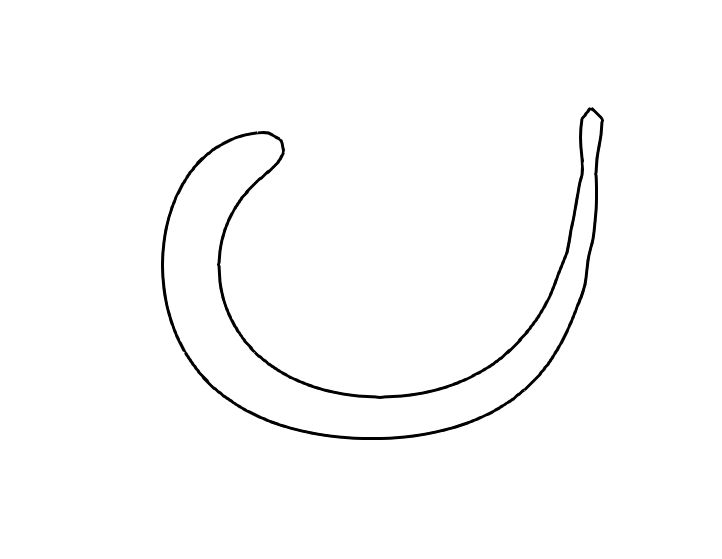}\\
        \includegraphics[width=\textwidth]{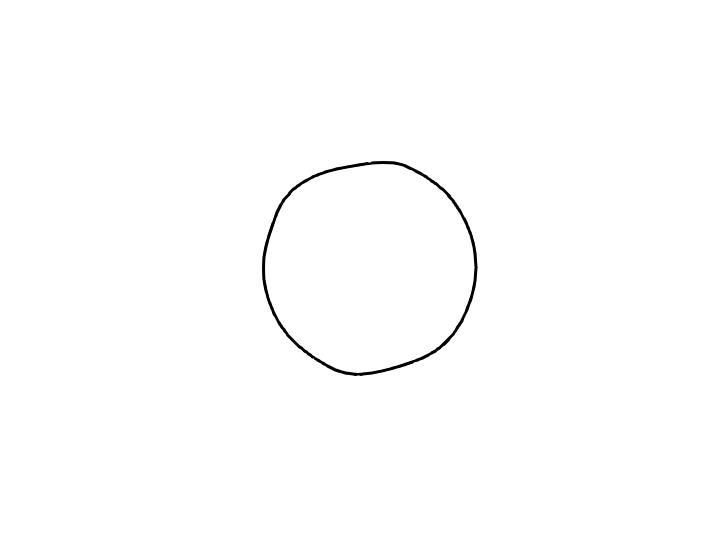}   
        \subcaption{Isocontour of drop with interIsoFoam for res.~64 at: top $t=0$\second, middle $t=1.5$\second, bottom $t=3$\second}
        \label{fig:reversedVortexFlow2Dc}
    \end{subfigure}
    \caption{Error visualization for 2D sphere in reversed vortex flow for different resolutions.}
\end{figure}
\Cref{fig:reversedVortexFlow2Da} shows the absolute shape error. We see a similar error distribution for different resolutions and solvers as before. Basilisk and interIsoFoam exhibit the smallest error, while interFoam with its algebraic transport accumulates the largest errors. The visualization of this error is shown on the right in \cref{fig:reversedVortexFlow2Db,fig:reversedVortexFlow2Dc}. The strong contour deformation exposes the reason for the shape error. 
In this case, interFoam achieves lower $L_1$ errors as for the diagonal flow. This can be explained by the fact that for the vortex flow the velocity aligns with normals of at least some cell faces. This is favorable for the algebraic transport. However, for the diagonal flow, there is no such alignment. The improved performance of Fluent for the vortex flow is difficult to explain without further details on the interface transport.


\subsubsection{3D sphere in vortex flow}
Extending the test of the previous paragraph to three dimensions, we initialize a sphere of radius $R=0.15$\meter at $(x_0,y_0,z_0)=(0.35,0.35,0.35)$\meter in a domain of 1\texttimes{}1\texttimes{}1\cubicmeter. The velocity is given by
\begin{align}
u & =2\sin^2(\pi x)\sin(2\pi y) \sin(2\pi z) \cos\left( \pi t /t_{end}\right)\\
v & =-\sin^2(\pi y)\sin(2\pi x) \sin(2\pi z) \cos\left( \pi t /t_{end}\right)\\
w & =-\sin^2(\pi z)\sin(2\pi x)\sin(2\pi y) \cos\left( \pi t /t_{end}\right) \,.
\end{align}
The resolution starts at 64\texttimes{}64\texttimes{}64 with two levels of refinement and an end time of $t_{end}=3$\second.
\begin{figure}[tb]
    \centering
    \begin{subfigure}{0.5\textwidth}
        \includegraphics[width=\textwidth]{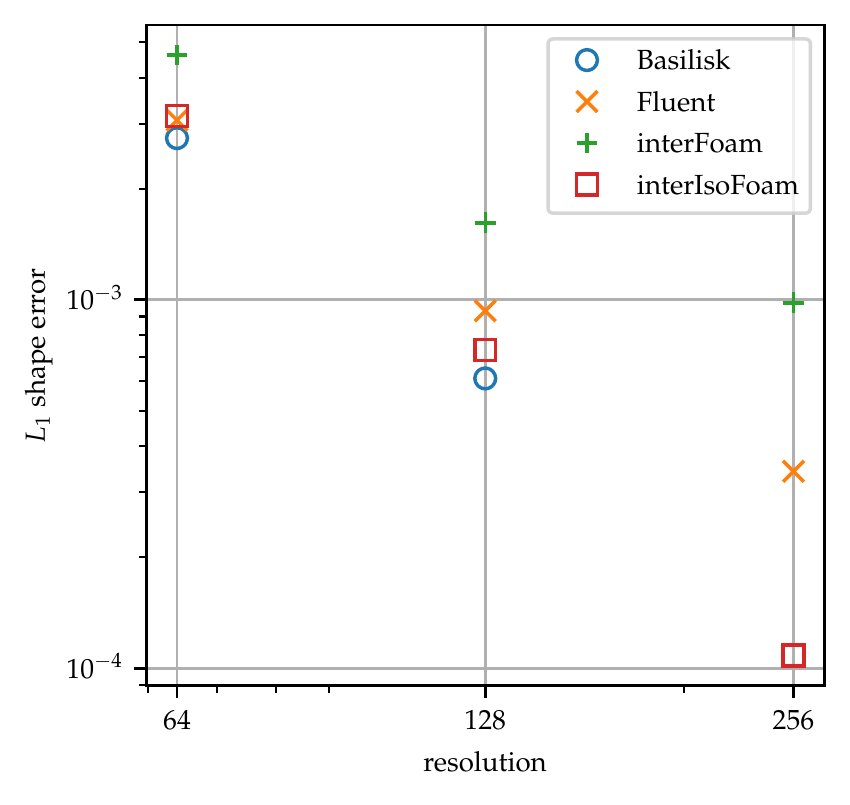}
        \subcaption{$L_1$ shape error for 3D sphere in reversed vortex flow and different resolutions.}       
        \label{fig:reversedVortexFlow3Da}
    \end{subfigure}
    \hspace{0.5cm}
    \begin{subfigure}{0.2\textwidth}
        \centering
        \includegraphics[width=\textwidth]{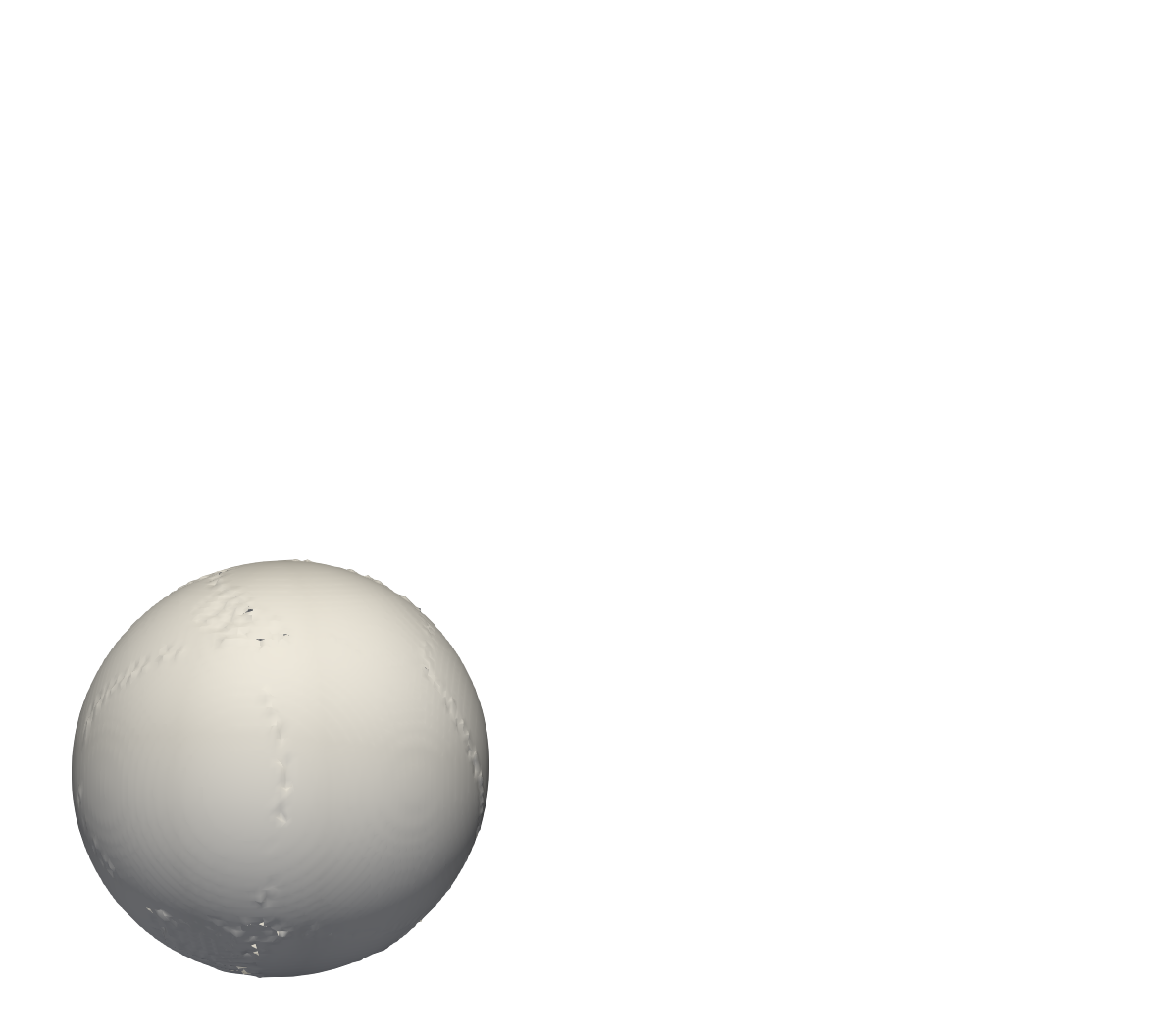}
        \includegraphics[width=\textwidth]{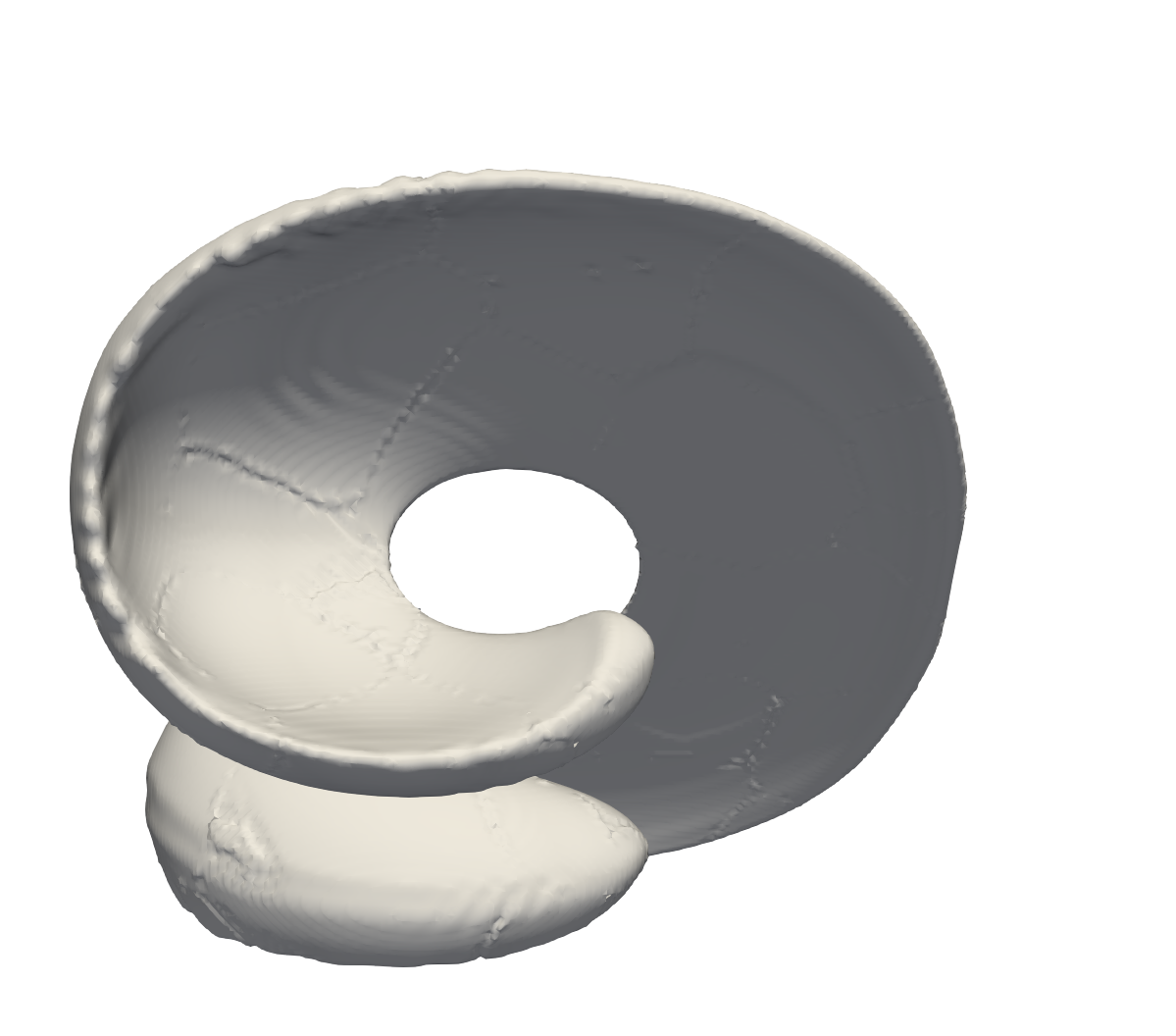}
        \includegraphics[width=\textwidth]{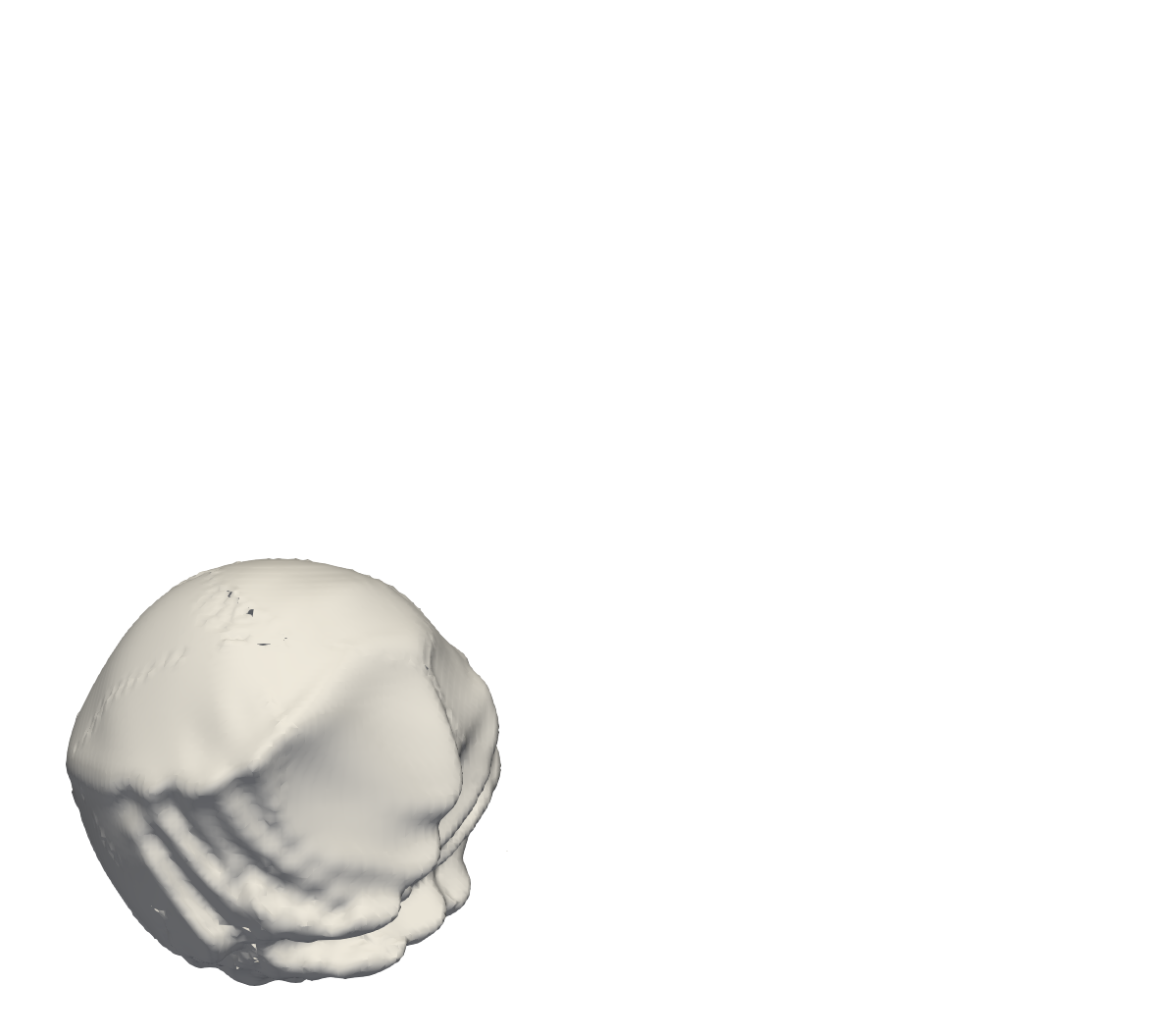}
        \subcaption{Isocontour of drop with interFoam for res.~256 at: top $t=0$\second, middle $t=1.5$\second, bottom $t=3$\second .}
        \label{fig:reversedVortexFlow3Db}
    \end{subfigure}
    \hspace{0.5cm}
    \begin{subfigure}{0.2\textwidth}
        \centering
        \includegraphics[width=\textwidth]{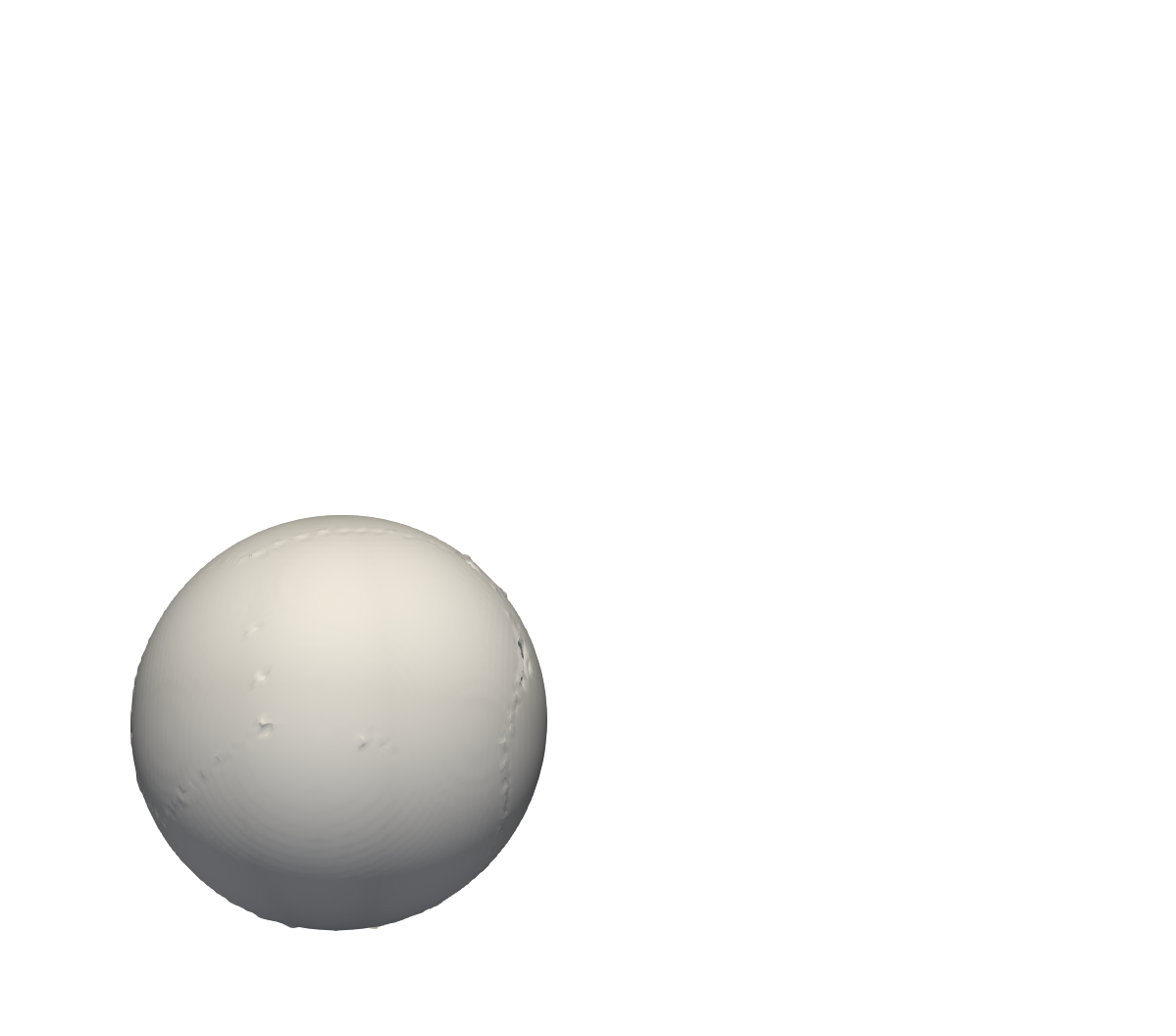}
        \includegraphics[width=\textwidth]{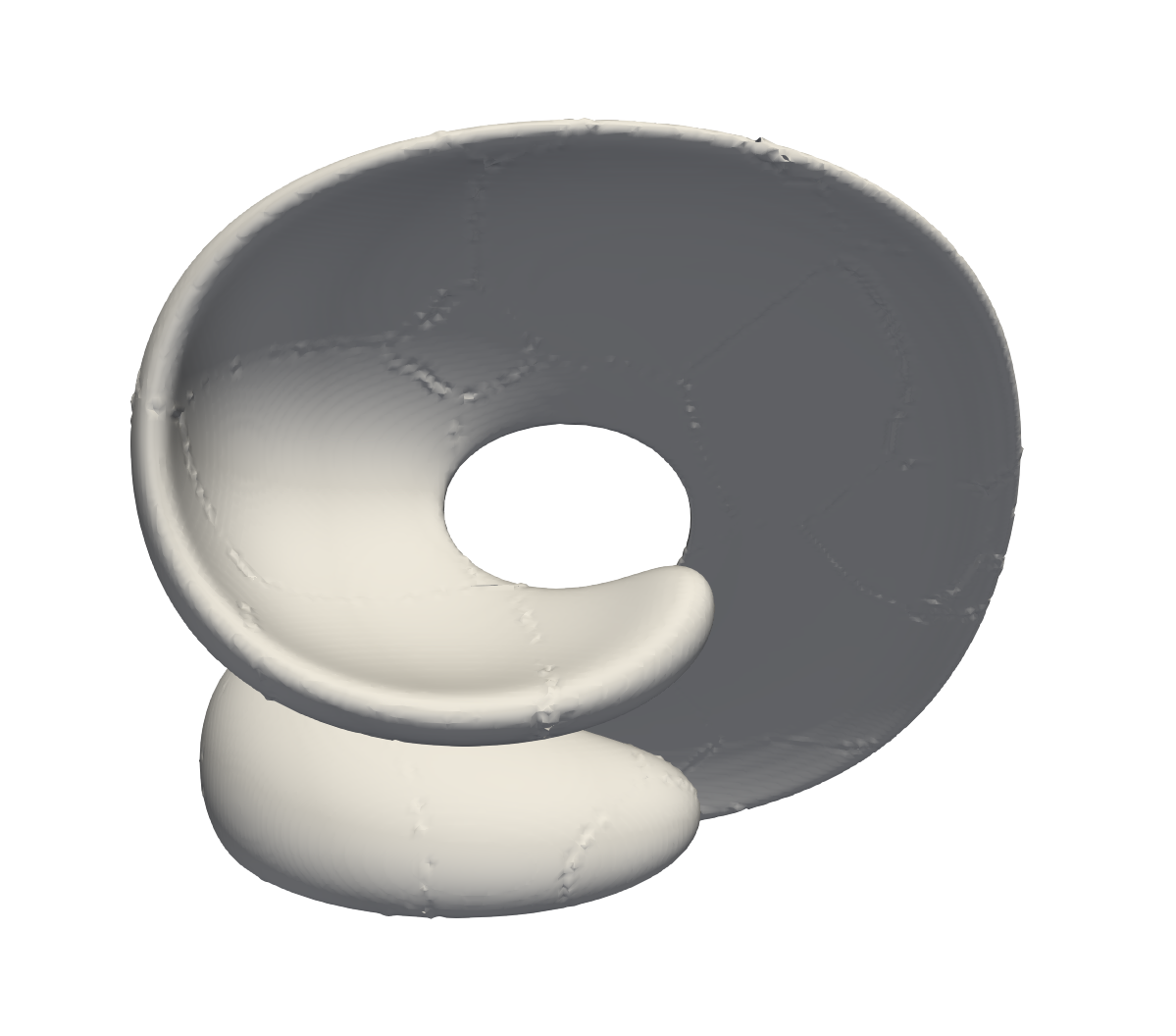}
        \includegraphics[width=\textwidth]{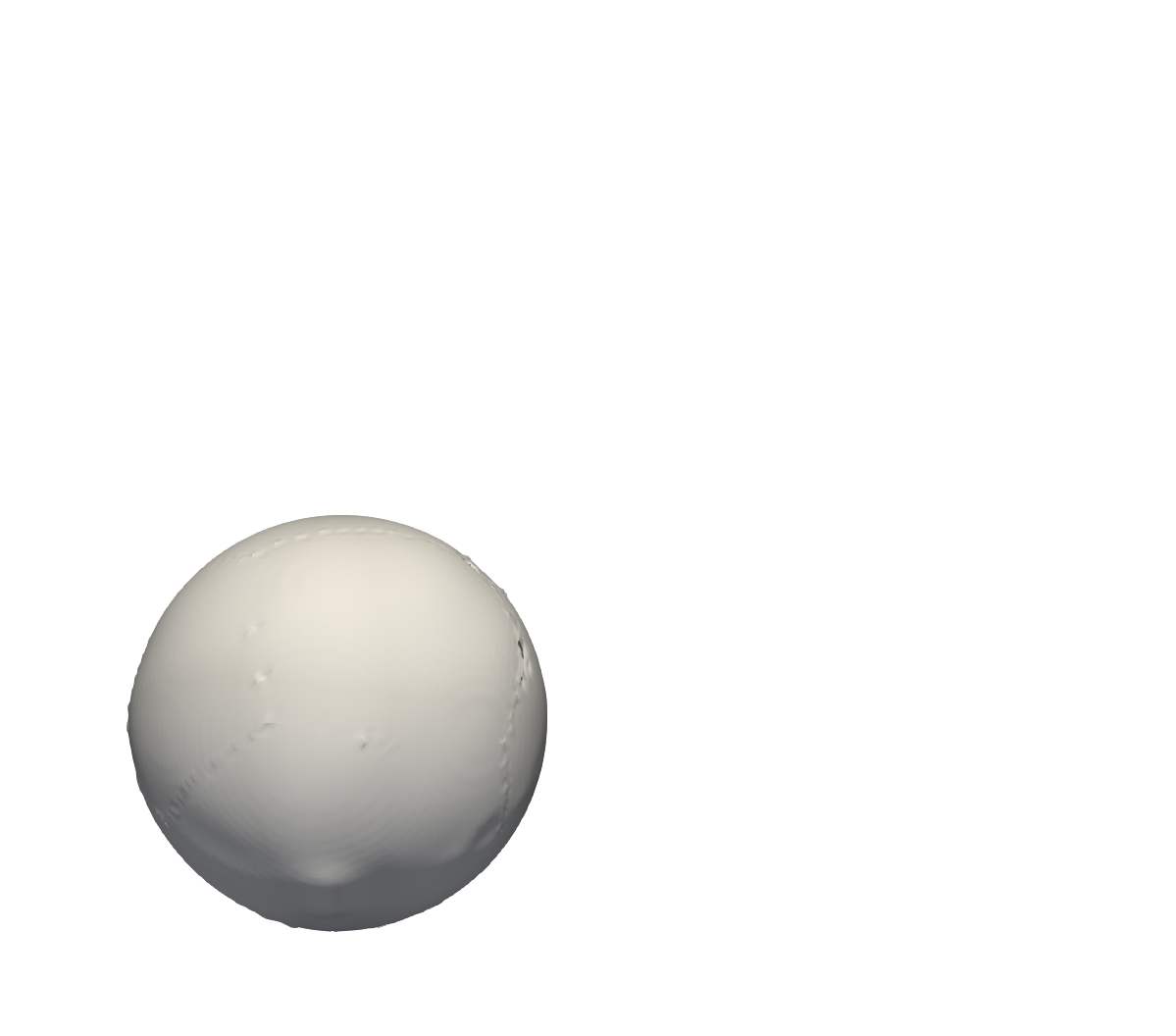}
        \subcaption{Isocontour of drop with interIsoFoam for res.~256 at: top $t=0$\second, middle $t=1.5$\second, bottom $t=3$\second.}
        \label{fig:reversedVortexFlow3Dc}
    \end{subfigure}
    \caption{$L_1$ shape error for 3D sphere in reversed vortex flow for different resolutions.}
\end{figure}

\Cref{fig:reversedVortexFlow3Da} shows that for a similar sized domain and drop, the error distribution between the solvers and resolutions stays consistent. Only Fluent appears to improve in three dimensions but still the error remains between the algebraic advection solver interFoam and the other geometric transport solvers Basilisk and interIsoFoam. Furthermore, the errors for all resolutions move closer together. Considering that the accuracy of algebraic transport depends on the alignment between flow
direction and cell face normals, this is not surprising as this alignment depends on cell topology not on spatial
resolution. However, looking at \cref{fig:reversedVortexFlow3Db,fig:reversedVortexFlow3Dc}, the deviation from the initial sphere shape at the end of the simulation is clearly visible with interFoam, even for the finest resolution of 256~cells. 


%% file: sections/results_hydrodyn.tex
\subsection{Hydrodynamic cases}
\label{sec:benchmark-hdyC}
The hydrodynamic cases quantify the numerical accuracy of the physical interaction between convection, diffusion and balancing surface tension effects with pressure difference. Three fluid pairings are considered. Their fluid properties can be found in \cref{tab:fluidprops}. These span a density ratio of 1.6 to 838.8 and a kinematic viscosity ratio of 0.06 to 15.3. A total of four cases is investigated, some of them in 2D and 3D, namely stationary droplet, translating droplet, oscillating droplet and oscillating wave.
\begin{table}[ht]
    \footnotesize
    \begin{tabular}{ lp{13ex}p{17ex}p{17ex}p{18ex} }
         \toprule
         @20\degreec & density $\rho$\newline (\negativeunitspace\kilogramspercubicmeter)  & kin. viscosity $\nu$\newline (\negativeunitspace\squaremeterpersecond) & dyn. viscosity $\mu$\newline (\negativeunitspace\pascalsecond)  & surface tension $\sigma$\newline (\negativeunitspace\newtonpermeter)  \\
         \midrule
         water\cite{Lemmon2022, VDI2010} & 998.2 & 1e-6 & 9.982e-4 & \multirow{2}{10em}{72.74e-3\cite{IAPWS2014}} \\
         air\cite{VDI2010} & 1.19 & 1.53e-5 & 1.8207e-5 &  \\
         \midrule[0.2\lightrulewidth]
         oil (Ravenol CLP 220~\cite{Ravenol2022}) & 888  & 2.71e-4\footnote{extrapolated value} & 0.240648 & \multirow{2}{10em}{32.9e-3\footnote{value for NACA reference oil from~\cite{Ross1950}}} \\ 
         air\cite{VDI2010} & 1.19 & 1.53e-5 & 1.8207e-5 &  \\
         \midrule[0.2\lightrulewidth]
          oil (Novec 7500~\cite{Novec2022}) & 1614 & 0.77e-6 & 1.24278e-3 & \multirow{2}{10em}{49.5e-3\cite{Brosseau2014}} \\
          water\cite{Lemmon2022, VDI2010} & 998.2 & 1e-6 & 9.982e-4 &   \\ 
        \bottomrule
    \end{tabular}
    \caption{Fluid pairings under consideration for hydrodynamic test cases.}
    \label{tab:fluidprops}
\end{table}
The time step for each case is restricted by the CFL number as well as the capillary time step constraint 
\begin{equation}
    \Delta t_\sigma = \sqrt{\frac{(\rho^- + \rho^+)\Delta x^3}{2\pi\sigma}}\,,
    \label{eq:capillary-time-step}
\end{equation}
as found in \cite{DENNER201524} for the respective resolutions~$\Delta x$.

For the stationary and translating droplet test cases (both 2D and 3D), the following error metrics considering velocity are examined:
\begin{align}
    L_1(\v) = & \frac{1}{N_\text{cells}}\sum_i^{N_\text{cells}}|\v_i - \v_\text{ref}|,
    \label{eq:l1error} \\
    L_\infty(\v) = & \max(|\v_i - \v_\text{ref}|),
    \label{eq:maxerror}
\end{align}
with $\v_\text{ref}$ denoting the uniform background velocity, being zero for the stationary droplet.


%
%

\subsection{Volume fraction initialization}
\label{sec:vof-init}
\begin{figure}[t]
    \centering
    \includegraphics[width=0.3\textwidth]{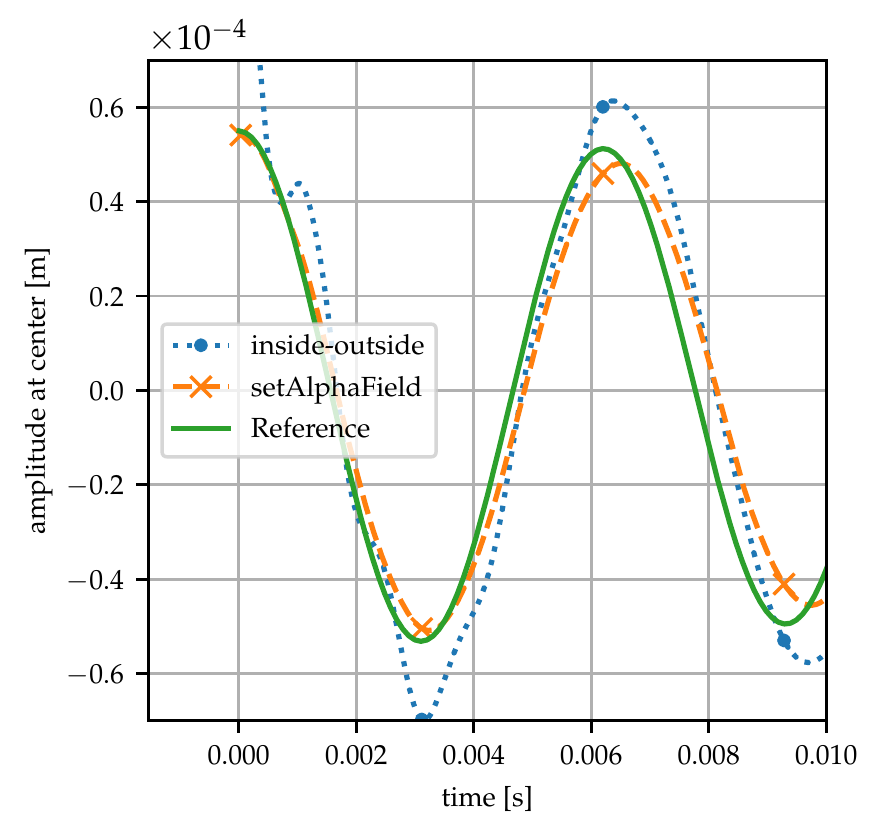}
    \includegraphics[width=0.3\textwidth]{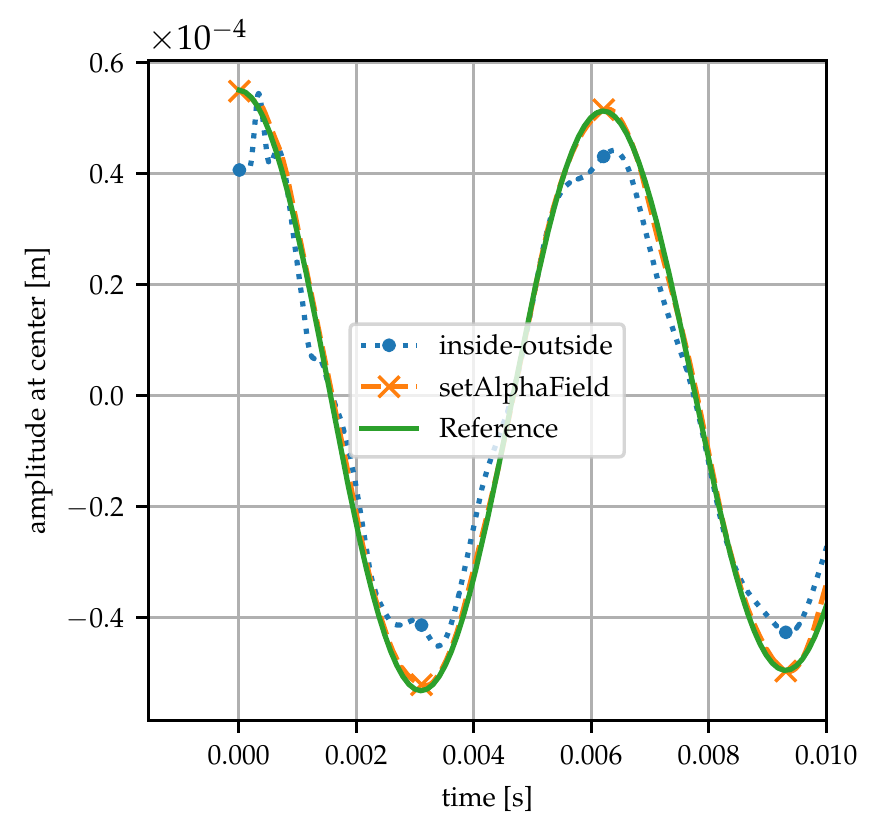}
    \includegraphics[width=0.3\textwidth]{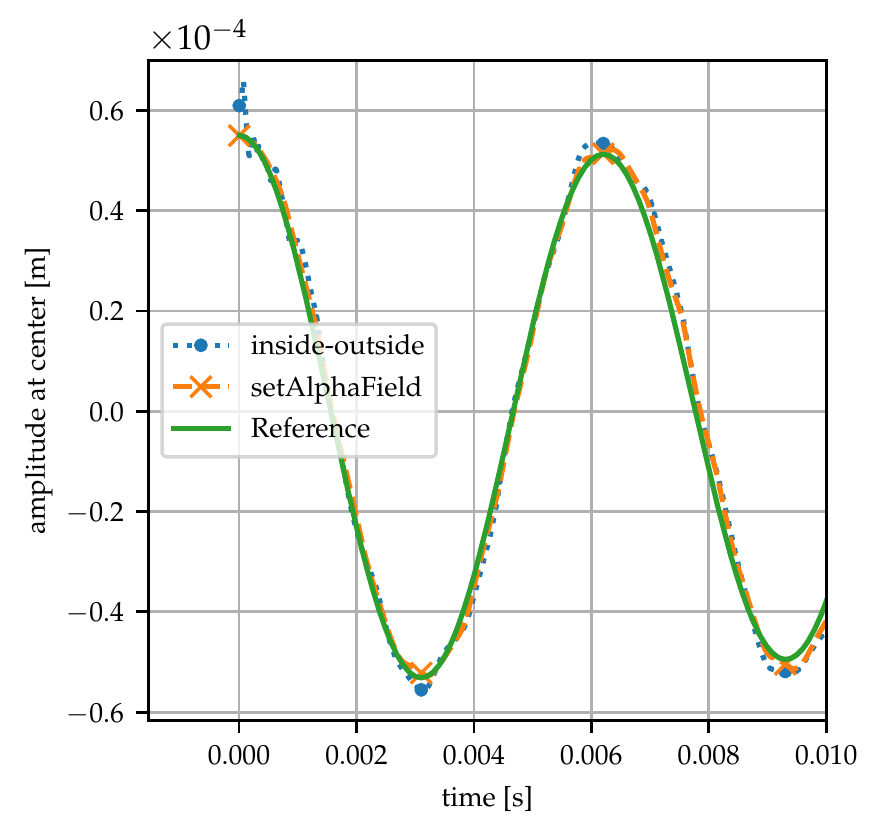}    
    \caption{Influence of different initialization algorithms for the 2D oscillating wave for three different resolutions, from left to right: 32, 64, 128. The solver interIsoFoam has been used for these computations with the water/air fluid pairing.}
    \label{fig:init_comp}
\end{figure}
Before continuing the discussion on hydrodynamic benchmark results, we want to highlight the influence of initialization accuracy here.
In the previous \cref{sec:benchmark-convC} this was less of an issue since the flow fields are all reversed and the initial volume fraction field also serves as the reference. So while the initial drop shape may differ between solvers, the error-measurement only compares their drop shape to itself.
For the hydrodynamic cases, this is not valid anymore. Observe the solution quality difference, especially for coarse resolutions, in \cref{fig:init_comp} which shows two options available in OpenFOAM. With the "inside-outside"
approach, the volume fraction value of a cell is determined solely by evaluating \cref{eq:phaseind} for its centroid $\x_c$.
This means $\alpha_c$ assumes values of either 1 or 0, depending on whether $\x_c \in \Omega^+$ or $\x_c \in \Omega^-$.
Consequently, this approach introduces rather large errors in the volume fraction.
The more accurate approach, as used in \texttt{setAlphaField}, determines the actual overlap between initialized shape contour and grid cells and by doing so calculates the volume fraction. The error decreases with increasing resolution, but must be kept in mind when comparing different solvers. OpenFOAM, TwoPhaseFlow and Basilisk all offer similar initialization algorithms. 
For Fluent, there are different options to define the phase distribution (here, custom field functions are used). However, phase initialization is based on the "inside-out" method only, albeit with a default "smoothing" step controlled by a relaxation parameter. The smoothing step adds one or more diffusion steps to the starting form of the "inside-out" volume fraction field~\cite{ansysfluenttheory2020r1}. Throughout this work, the default of one smoothing step with a factor of 0.25 is used.

\subsubsection{2D stationary droplet}
\label{sec:2dstationaryD}
\input{figures/latex-included/stationaryDrop2D_mean}%
\input{figures/latex-included/stationaryDrop2D_maxError}%
\input{figures/latex-included/stationaryDrop2D_violinPlots}%
To analyze so-called parasitic currents in 2D, a drop with $R=10^{-3}$\meter is initialized at the center of a Cartesian mesh of 32\texttimes{}32 cells in a rectangular domain of 0.01\texttimes{}0.01\squaremeter. The end time is set to $t_\text{end}=0.1$\second. Two additional levels of refinement are investigated. The error evaluation is done using \cref{eq:l1error,eq:maxerror}. Plotting the mean and max error over time for the middle resolution for all fluid pairings reveals the differing behavior of solvers over time. In general, Basilisk and interFlow, using plicRDF for curvature (abbr. interFlow-RDF), produce the smallest error for all fluid pairings. As expected, the error magnitude decreases across-the-board for higher viscous fluids, see Ravenol/air vs. water/air. The fluid pairing Novec/water with a density and viscosity ratio of close to one reduces the error compared to water/air by about one order of magnitude. Furthermore, the max error (\cref{fig:max_err_2Dstat}) exceeds the mean error (\cref{fig:mean_err_2Dstat}) by at least one order of magnitude. In addition, the plots reveal that only for Basilisk and interFlow-RDF the error reduces over time. None of the others are unstable, but reach some kind of plateau state.%

To capture the essence of the simulations over all resolutions, fluid pairings and their time-dependent changes, we introduce in \cref{fig:violinPlots_2DstationaryD} violin plots which include for each solver and resolution the mean (darker color) and max error (lighter color) with their respective max and min values as well as their time-variant distribution.  For these distributions, only the values from the second half of simulated time, $t > \frac{1}{2}t_\text{max}$, are considered to reduce the
influence of initialization, shown in \cref{sec:vof-init}.
This overview reveals a multitude of things, the most obvious is a missing trend of a reduction in error with increased resolution, meaning the error for a resolution 128\texttimes{}128 might be higher than for 64\texttimes{}64. This stands in strong contrast to the previous findings regarding the convection error in \cref{sec:benchmark-convC}.


\subsubsection{3D stationary droplet}
\label{sec:3dstationaryD}
Extending the above described setup to three dimension in a cubic domain of
0.01\texttimes{}0.01\texttimes{}0.01\cubicmeter and a droplet radius of $R=10^{-3}$\meter, the deviation from the initialized drop at rest is studied. The drop is initialized at the center of the domain. Errors are once more evaluated via \cref{eq:l1error,eq:maxerror}.
The end time is set to $t_\text{end}=0.1$\second with the coarsest resolution being 32\texttimes{}32\texttimes{}32 followed by two refinement steps.
\Cref{fig:violinPlots_3DstationaryD} shows again the missing correlation of error magnitude and resolution. For instance, considering the data of Fluent, for water/air the mean and max error increase for increased resolution. In a Ravenol/air setup, however, the mean error decreases while the max error still increases. For interIsoFoam, the trend is completely reversed, decreasing errors for water-air, but increasing errors for Ravenol-air. In general, an increasing error for finer refinement appears counter-intuitive but can be explained by the less diffused interface and, hence, higher more localized error. This primarily hits the geometrical solvers who base their curvature on the second derivative of the $\alpha$-field, such as Fluent and interIsoFoam. Basilisk's poor performance for Ravenol-air for 128\texttimes{}128\texttimes{}128 cannot be explained by this, but should be noted here, as it stands in harsh contrast to the 2D case in \cref{sec:2dstationaryD}.
Overall, the errors are close in magnitude, with Basilisk still outperforming the unstructured solvers, but less distinct than in two dimensions (see \cref{fig:violinPlots_2DstationaryD} for reference). In general, compared to the 2D case, the mean and max errors differ more strongly, covering at least one order of magnitude, often even two.
\input{figures/latex-included/stationaryDrop3D_violinPlots}


\subsubsection{2D translating droplet}
\label{sec:2dtranslatingD}
The following case upgrades the 2D stationary droplet case, see \cref{sec:2dstationaryD},  with a background velocity field of $\mathbf{U}=(1.0,0)$\meterpersecond to evaluate their error superposed with the advection errors investigated in \cref{sec:benchmark-convC}.
For this purpose, we initialize a drop with $R=10^{-3}$\meter at $(2R,5R)$ in a rectangular domain of 0.02\texttimes{}0.01\squaremeter
and simulate up to an end time of $t_\text{end}=0.015$\second. The domain is resolved with 64\texttimes{}32 and subsequently
twice refined for this study.

\input{figures/latex-included/translatingDrop2D_violinPlots}
\Cref{fig:violinPlots_2DtranslatingD} summarizes the different solver performances for varied resolution and fluid pairings. The evaluation is done using \cref{eq:l1error} and \cref{eq:maxerror} with $\v_\text{ref}=\mathbf{U}$.
Focusing first on the right hand side of \cref{fig:violinPlots_2DtranslatingD}, the OpenFOAM based solvers interFlow, interFoam and interIsoFoam, perform more similarly compared to Fluent and Basilisk with their max and mean error being of a comparable order. Only interFoam and interIsoFoam exhibit outliers for the max error when it comes to the fluid pairing of Novec and water, even though the density ratio is nearly one here.
For all fluid-pairings and resolutions, Basilisk shows the best results. Fluent stays between Basilisk and the OpenFOAM solvers. This is surprising based on the 2D convection and 2D stationary droplet case, where both errors exceeded for instance interFlow-RDF by one order of magnitude. Due to the closed source and commercial nature of Fluent, we can only assume that the errors either cancel each other for this specific case or some additional unknown feature is at work.


\subsubsection{3D translating droplet}
Switching to a cuboid domain of 0.02\texttimes{}0.01\texttimes{}0.01\cubicmeter, the drop is initialized with $R=10^{-3}$\meter at $(2R,5R,5R)$. The translation velocity is as in the 2D case chosen as $\mathbf{U}=(1.0,0,0)$\meterpersecond while translating to an end time of $t_\text{end}=0.015$\second. The domain is resolved with 64\texttimes{}32\texttimes{}32 and twice refined to study the influence of grid resolution.\\
\input{figures/latex-included/translatingDrop3D_violinPlots}
Comparing first the mean errors for different fluid pairings in 3D as visible in \cref{fig:violinPlots_3DtranslatingD} with the 2D errors in \cref{fig:violinPlots_2DtranslatingD}: The errors stay in a similar range, even slightly improving in three dimensions. Especially Fluent and interFlow-RDF improve.
Across the board, the solvers perform worst for water/air as expected for low viscosities,
highest density ratio and largest surface tension coefficient. It is also worth mentioning that while the mean error decreases for most solvers with increasing resolution, the maximum error often increases. This effect was less noticeable for the stationary droplet 3D, see \cref{fig:violinPlots_3DstationaryD} and must stem from the additional phase transport. Different phase/interface positions for each time step might result in disadvantageous normal calculations for the geometrical advection schemes which introduce additional errors. However, even while max errors increase, the mean errors stay in a similar range to the 3D stationary droplet or partly improve.


\subsubsection{2D oscillating wave}
The capillary wave with an initial interface position $h(x,0)= a\cos{(2\pi/L(x-L/2))}+L/2$ with $a=\expo{5.5}{-5}$\meter and $L=\expo{2.6}{-3}$\meter is initialized in domain of L\texttimes{}L. Gravitational effects are not considered. The domain is resolved 32\texttimes{}32 and refined twice. The considered physical time span is such that 5 oscillations are tracked.
The interface height at the center of the domain is tracked and compared to the analytical solution by Prosperetti (eqs. (24) and (25) in \cite{Prosperetti1981}).
Since the high viscosity of Ravenol causes overdamping, see \cref{tab:capillary_wave_reference} (frequency close to zero), only the fluid pairings water/air and Novec/water are discussed.
To summarize the temporal evolution of the difference between the numerically predicted
interface height and the asymptotic reference solution, the time-averaged
relative amplitude error
\begin{equation}
    \Bar{\epsilon}_{a,rel} = \frac{\Delta t}{\tau}
        \sum_i \frac{|a_{i,\text{num}} - a_{i,\text{ref}}|}{a_0}
    \label{eq:timeamplitudeerror}
\end{equation}
is evaluated. Here, $\tau$ denotes the total physical time simulated, the subscript $i$ refers
to the $i$-th time step, $a_{i,\text{num}}$ and $a_{i,\text{ref}}$ refer to the numerically
approximated interface height and the the interface height according to the reference 
at time step $i$ and
$a_0 = \expo{5.5}{-5}$\meter denotes the initial amplitude.

\begin{table}[ht]
    \centering
    \begin{tabular}{lll}
    \toprule
     fluid pairing  & decay rate $\delta$ [1/s] & frequency $f$ [1/s]\\
     \midrule
       water/air & 11.46 & 161.13 \\
       Novec/water & 18.90 & 79.53 \\
       Ravenol/air & 181.05 & 0.001 \\ 
     \bottomrule
    \end{tabular}
    \caption{Parameters of the asymptotic solution $h(0,t)-L/2=a \exp(-\delta t) \cos(2\pi f t)$ to the capillary wave problem for~$t\to\infty$ by Prosperetti (eqs. (24) and (25) in \cite{Prosperetti1981}). The remaining parameters are given by $L=\expo{2.6}{-3}$\meter and $a=\expo{5.5}{-5}$\meter, respectively.}
    \label{tab:capillary_wave_reference}
\end{table}

\input{figures/latex-included/capillaryWave2D}
\input{figures/latex-included/oscillatingWave2D_accuPlots}

\Cref{fig:temporal_evolution_2DoscWave} displays the temporal evolution of the film height above the center for 
water/air and Novec/water at the finest resolution. Deviations are visible for interIsoFoam and Fluent. With interIsoFoam as
solver, the deacy of the amplitude
is slightly underestimated for water/air, a behaviour that is more pronounced for the Novec/water fluid pairing.
Results from Fluent
show a longer oscillation period and a faster amplitude decay than given by the reference solution for both
fluid pairings. The behaviour of Fluent can be explained, at least partially, by its rather inaccurate initialization
approach. As shown in \cref{sec:vof-init}, initialization errors impact the computed solution since they represent
deviations from the exact interface. All other solvers coincide with the reference solution for both fluid pairings.
These observations are also reflected in the errors evaluated according to \cref{eq:timeamplitudeerror}, displayed
in \cref{fig:time_averaged_2DoscWave}. Again, interIsoFoam and Fluent display the largest errors for both fluid pairings, with
the exception of water/air at the intermediate resolution. For that configuration, interIsoFoam yields the lowest error
among the benchmarked solvers. Given that the finest resolution underestimates the decay of the amplitude and the coarse
resolution tends to overestimate the decay, interIsoFoam probably obtains the best agreement with the reference for the
intermediate resolution. Besides interIsoFoam, all solvers show converging errors. Comparing fluid pairings,  the errors are
in the same order of magnitude for fixed resolution and solver.
A remark on the choice of boundary condition for the volume fraction field when using the
interFlow solver: with OpenFOAM's \verb+zeroGradient+ boundary condition, in principle a valid boundary condition for this case which has been used for interFoam and interIsoFoam, the results
showed no resemblance with the reference solution. However, when the
\verb+constantAlphaContactAngle+ boundary condition, provided by TwoPhaseFlow \cite{scheufler2021twophaseflow}, with a conatct angle of 90 degree, accuracy improved
distinctly as shown in \cref{fig:temporal_evolution_2DoscWave,fig:time_averaged_2DoscWave}.


\subsubsection{3D oscillating droplet}
As a test case in three dimensions with deforming interface an oscillating droplet is chosen. 
The droplet is initialized as a rotational ellipsoid
\begin{equation}
     \frac{x^2}{a_0^2} +\frac{y^2+z^2}{b_0^2}=1
\end{equation}
with semi-axes $a_0=\expo{5.25}{-4}$\meter and $b_0=\expo{4.8795}{-4}$\meter, respectively. The droplet's volume is equivalent to that of a spherical droplet of radius $R=\expo{0.5}{-3}$\meter. The droplet is placed at the center of a cubic domain of edge length $l=\expo{5}{-3}$\meter and the domain's center coincides with the origin. The domain is resolved with 50 cells in each direction and twice refined. Slip is prescribed as velocity boundary condition and a Neumann boundary condition for the pressure. Both fields are set to zero initially. A total physical time of $t=0.02$\second is simulated with a fixed time step determined by \cref{eq:capillary-time-step}.
For a droplet that is only slightly perturbed from a spherical shape, an analytical solution for the eigenshape at eigenfrequency $\omega_n$ has been given by \citet{lamb1975}. Rather than performing a surface harmonics expansion of the droplet and evaluating it over time to find the interface evolution, the simplified approach by \citet{shin2002} is followed. Thus, in order to compute the oscillation of the longest semi axis~$a(t)$, the amplitude of the lowest-order eigenshape ($n = 2$) is set to the initial semi axis length~$a_0$ minus the mean droplet radius~$R$. This leads to the reference solution 
\begin{equation}
    a(t) = R + \left(a_0-R\right) \exp\!{\left(-\frac{(n-1)(2n+1)\nu}{R^2}t\right)}  \cos(\omega_n t)~\text{for}~ n=2,
    \label{eq:lamb3D}
\end{equation}
where 
\begin{equation}
     \omega_n^2 =\frac{n(n+1)(n-1)(n+2)\sigma}{\left((n+1)\rho^- + n \rho^+\right)R_0^3}.
\end{equation}
Since \cref{eq:lamb3D}, by assumption of a viscous droplet surrounded by a fluid of comparatively negligible kinematic viscosity, is only valid for the fluid paring water/air, the kinematic viscosity~$\nu$ is set to the value for water (cf. \cref{tab:fluidprops}), i.e. $\nu=\nu^-$.

%
\Cref{fig:temporal_evolution_oscDrop3D} shows the temporal evolution of the initially longest semi-axis for 
water/air and Novec/water fluid pairings. Ravenol/air is not reported here as the high viscosity of the
oil prevents oscillatory behaviour similar to the oscillating wave. Note also that the reference solution is only plotted in \cref{fig:temporal_evolution_oscDrop3D}a and not b since Novec/water does not fulfill the criterion of a negligible ambient fluid viscosity. With water/air as fluid pairing, all
solvers match the oscillation period quite well. After $t=0.015$\second, small offsets become apparent,
for interIsoFoam and interFlow-RDF offsets are noticeable at earlier times already. Based on their previous behavior in other test scenarios, this is not explainable. \Cref{fig:accuPlots_3DoscD} highlights this discrepancy and even shows a increasing deviation with increasing resolution. For Fluent, the influence of volume fraction initialization is visible for around one and a half
oscillation periods. Afterwards its performance stabilizes and it performs rather well. This finding agrees with the significant improvement of Fluent's time averaged relative amplitude error with increasing resolution as visible in \cref{fig:accuPlots_3DoscD}, again pointing towards the drawback of inaccurate initialization. In general, all solvers overestimate the damping, probably due to realistic fluid pairing where the outer viscosity differs from zero. 
Regarding the fluid pairing Novec/water only the relative solver performance can be judged. Again, Fluent struggles with initialization which introduces additional harmonics. The dominant oscillating frequency lies in a similar range for all solvers, with interFoam and interIsoFoam as outliers. InterFlow-RDF overshoots massively towards one side. Assuming a good performance of Basilisk based on the previous results, only interFlow-fP shows a comparable behavior. 

\input{figures/latex-included/oscillatingDrop3D}
\input{figures/latex-included/oscillatingDrop3D_accuPlots}

%% file: figures/latex-included/stationaryDrop2D_mean.tex
\begin{figure}[tb]
    \centering
      \begin{subfigure}[b]{0.32\textwidth}
         \centering
         \includegraphics[width=\textwidth]{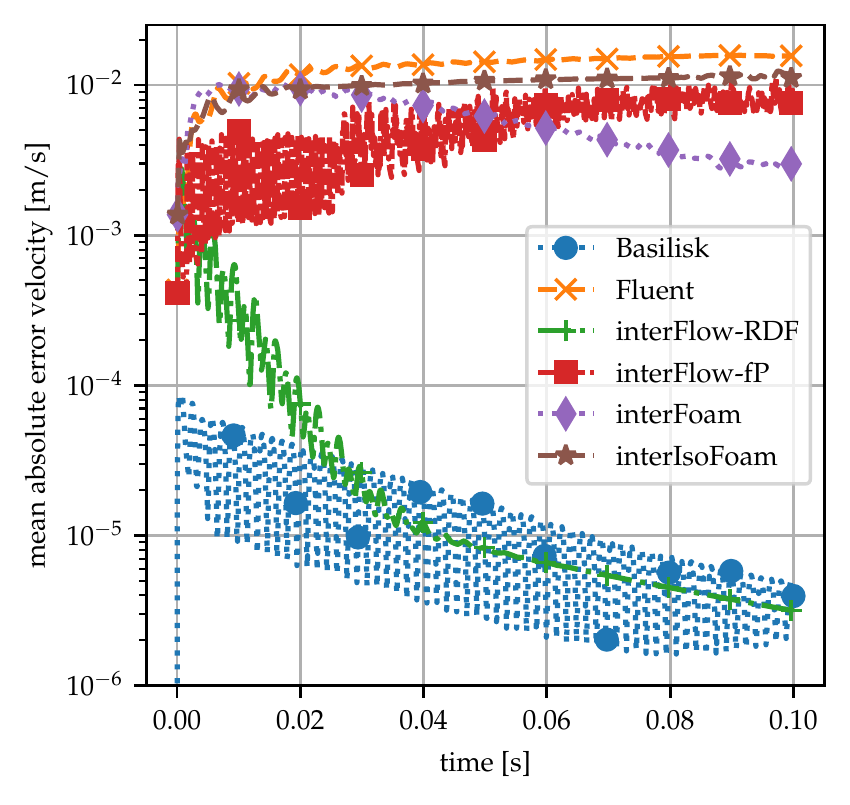}
         \caption{water/air, res: 64}
         \label{fig:stationaryDroplet2D_fluid_pairing__water-air_resolution__64__mean_absolute_error_velocity}
     \end{subfigure}
     \hfill    
     \begin{subfigure}[b]{0.32\textwidth}
         \centering
         \includegraphics[width=\textwidth]{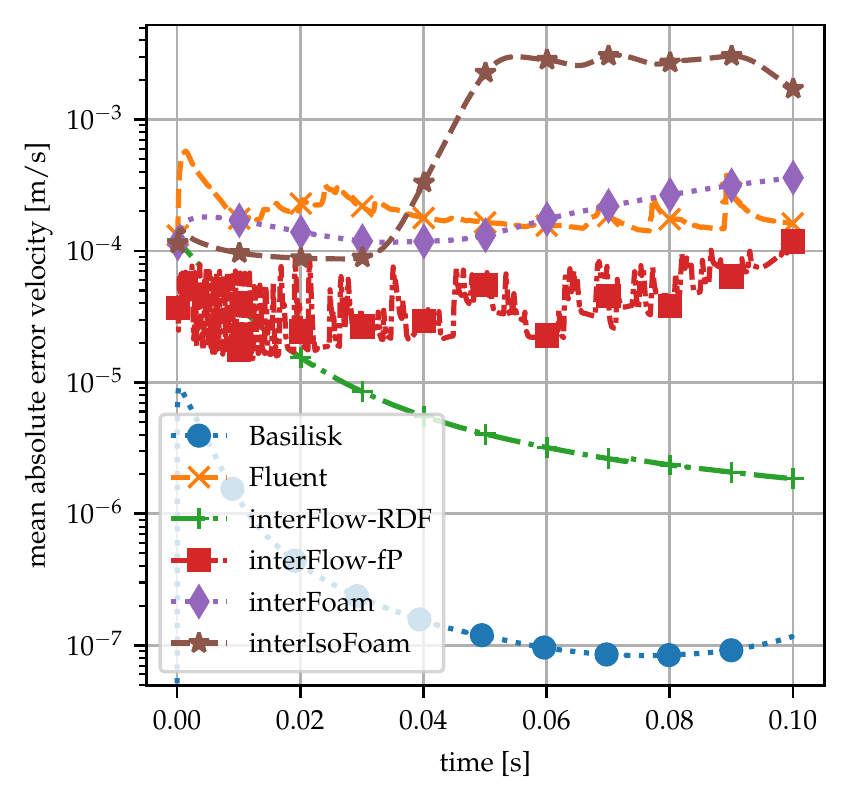}
         \caption{Ravenol/air, res: 64}
         \label{fig:stationaryDroplet2D_fluid_pairing__gearoil-air_resolution__64__mean_absolute_error_velocity}
     \end{subfigure}
     \hfill     
     \begin{subfigure}[b]{0.32\textwidth}
         \centering
         \includegraphics[width=\textwidth]{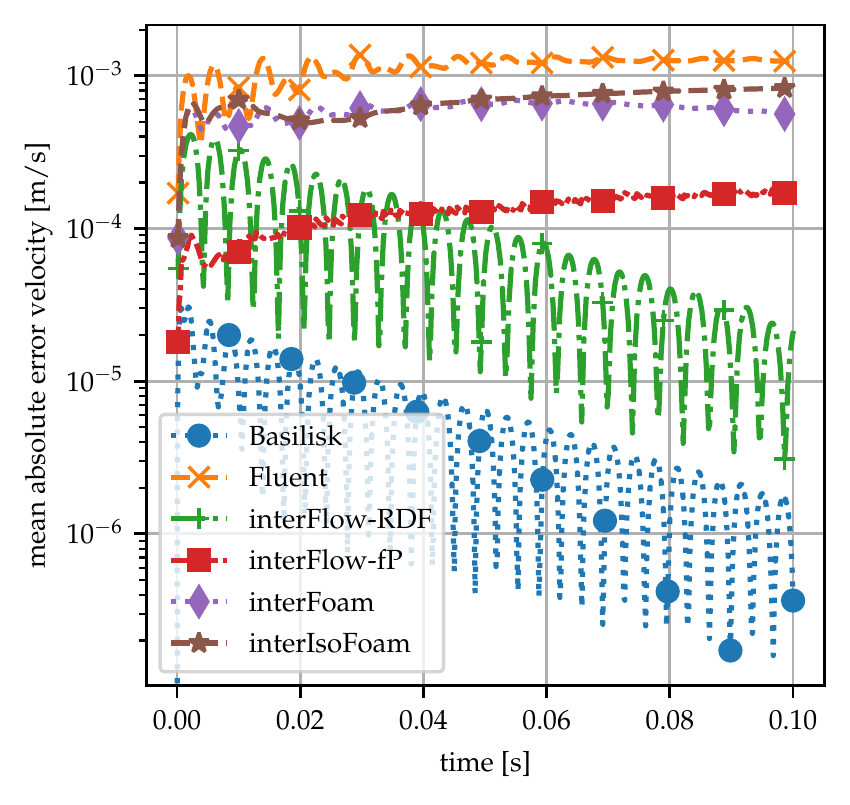}
         \caption{Novec/water, res: 64}
         \label{fig:stationaryDroplet2D_fluid_pairing__oil_novec7500-water_resolution__64__mean_absolute_error_velocity}
     \end{subfigure}     
     \newline
    \caption{Mean velocity error for 2D stationary droplet for different fluid pairings.}
    \label{fig:mean_err_2Dstat}
\end{figure}

%% file: figures/latex-included/stationaryDrop2D_maxError.tex
\begin{figure}[htb]
    \centering
      \begin{subfigure}[b]{0.32\textwidth}
         \centering
         \includegraphics[width=\textwidth]{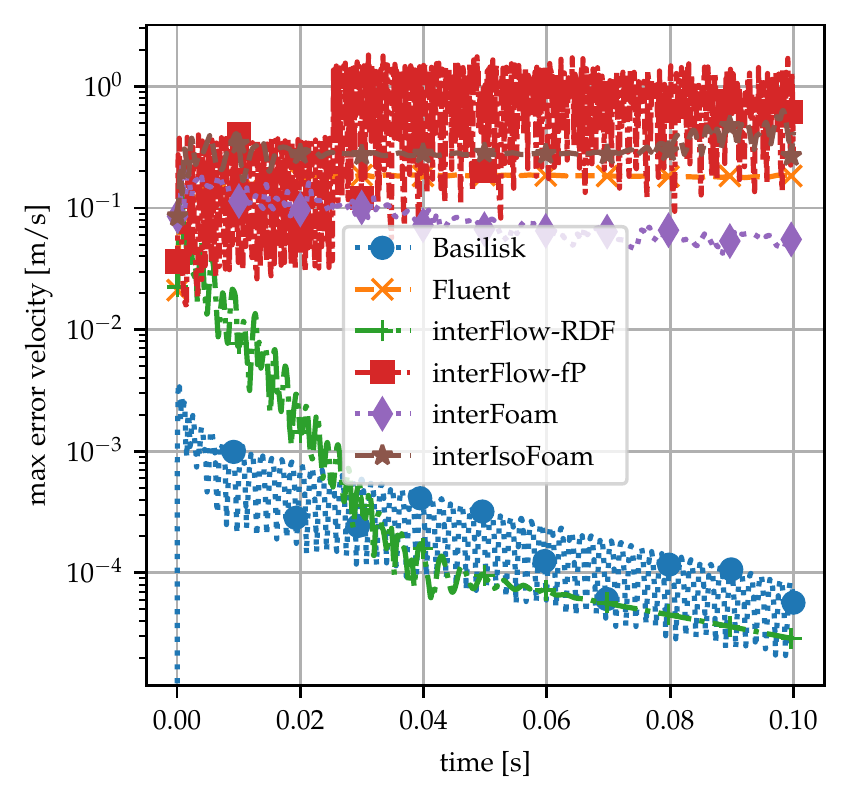}
         \caption{water/air, res: 64}
         \label{fig:stationaryDroplet2D_fluid_pairing__water-air_resolution__64__max_error_velocity}
     \end{subfigure}
     \hfill    
     \begin{subfigure}[b]{0.32\textwidth}
         \centering
         \includegraphics[width=\textwidth]{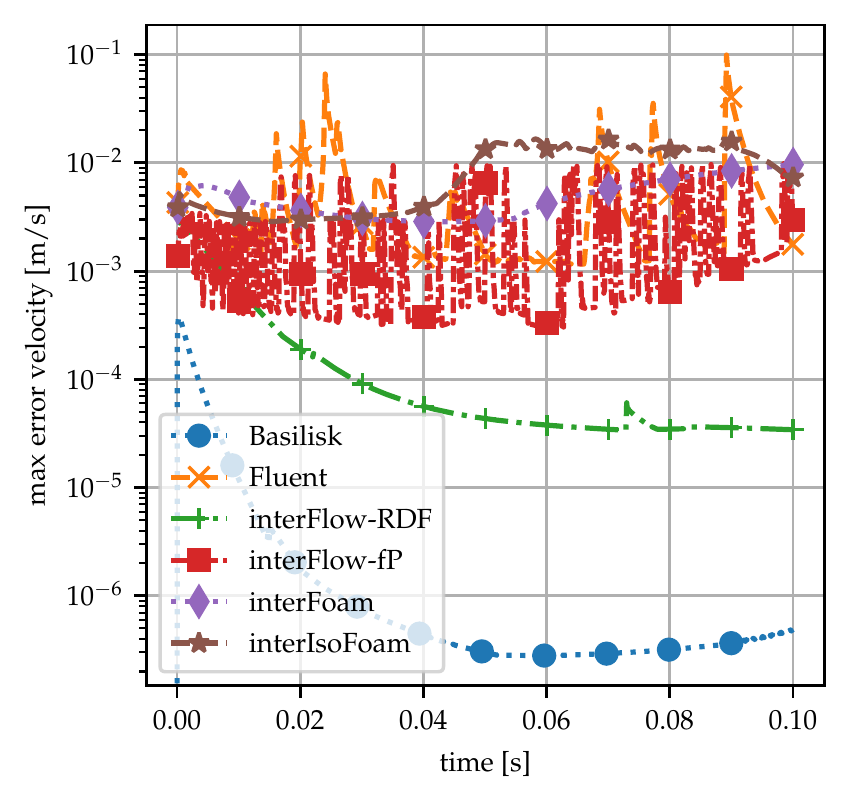}
         \caption{Ravenol/air, res: 64}
         \label{fig:stationaryDroplet2D_5}
     \end{subfigure}
     \hfill     
     \begin{subfigure}[b]{0.32\textwidth}
         \centering
         \includegraphics[width=\textwidth]{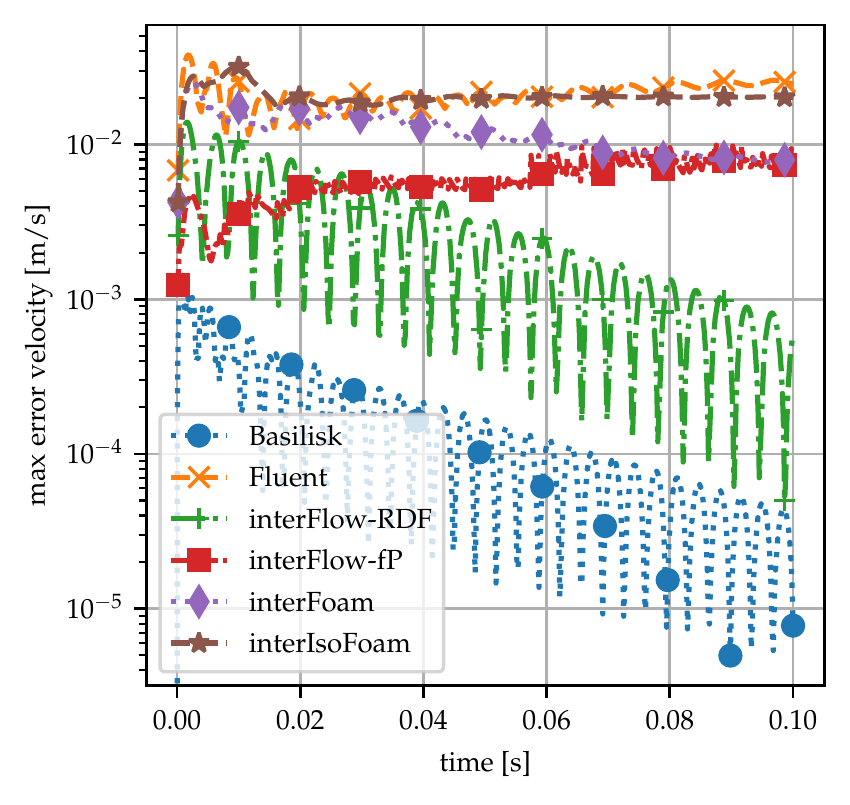}
         \caption{Novec/water, res: 64}
         \label{fig:stationaryDroplet2D_fluid_pairing__oil_novec7500-water_resolution__64__max_error_velocity}
     \end{subfigure}     
     \newline
    \caption{Max velocity error for 2D stationary droplet for different fluid pairings.}
    \label{fig:max_err_2Dstat}
\end{figure}

%% file: figures/latex-included/stationaryDrop2D_violinPlots.tex
\begin{figure}
    \centering
     \begin{subfigure}[b]{\textwidth}
        \includegraphics[width=\textwidth]{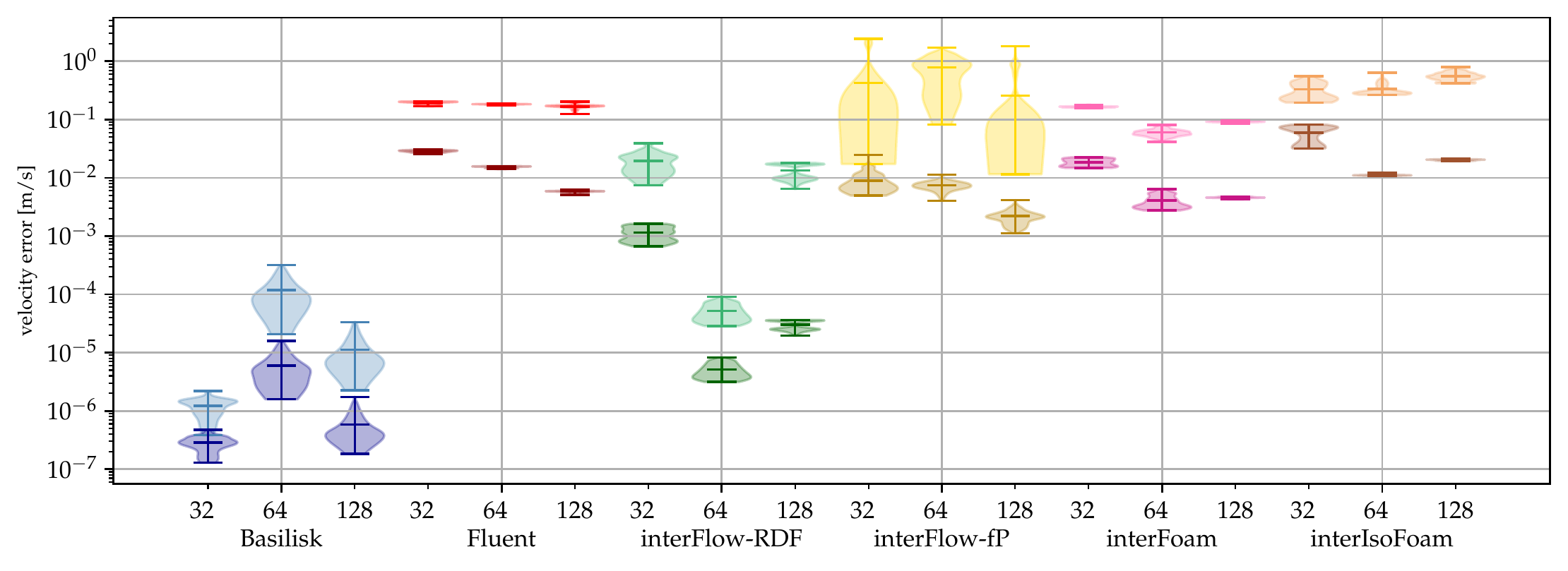}    
        \subcaption{water/air}
    \end{subfigure}
     \begin{subfigure}[b]{\textwidth}
        \includegraphics[width=\textwidth]{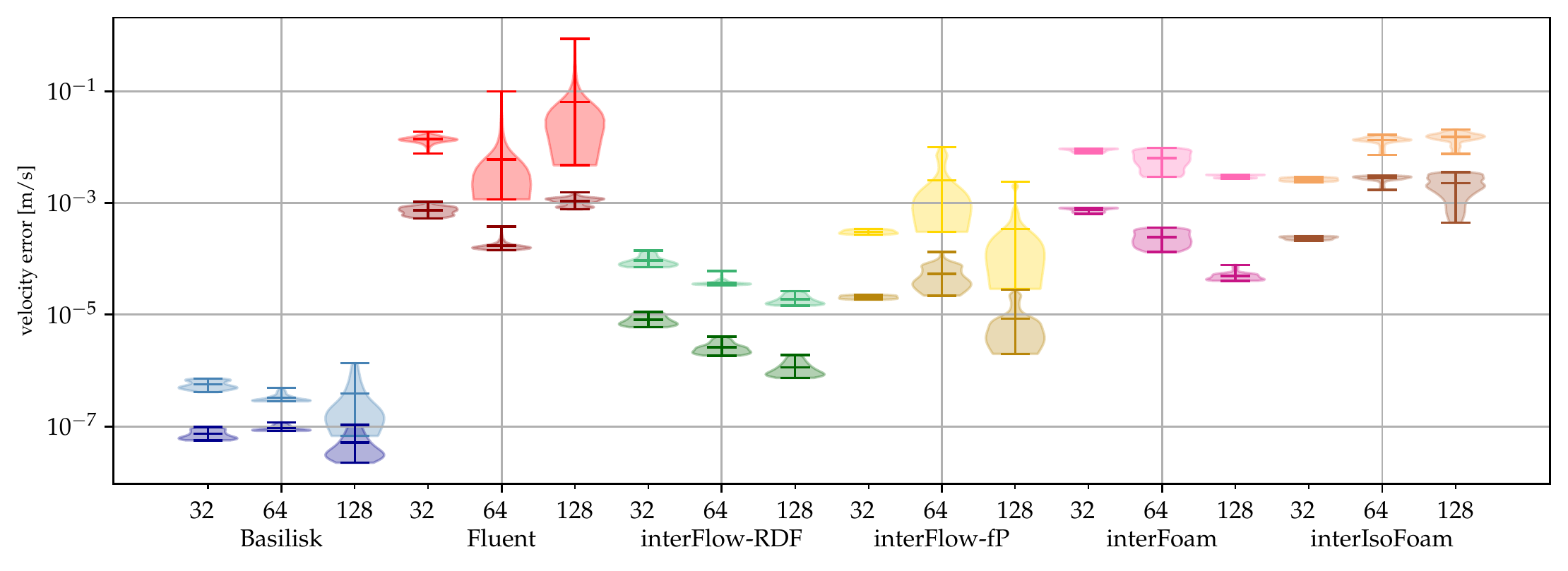}
        \subcaption{Ravenol/air}        
    \end{subfigure}
     \begin{subfigure}[b]{\textwidth}
        \includegraphics[width=\textwidth]{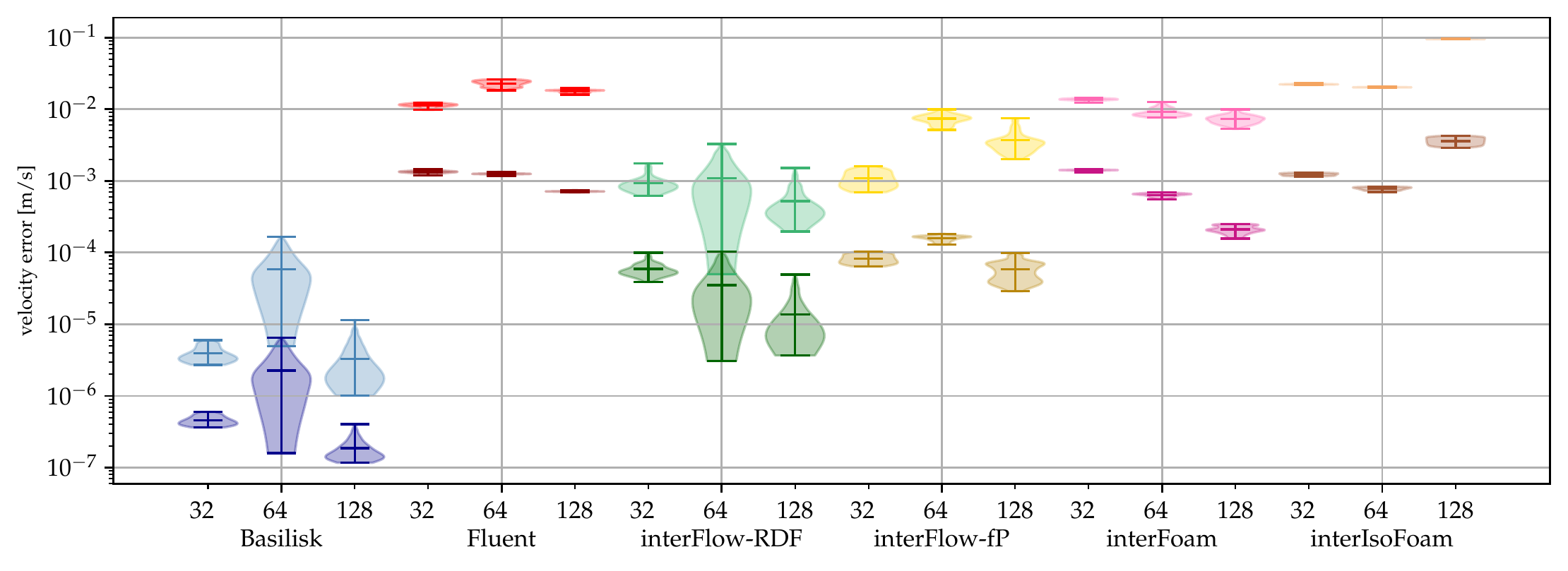}
        \subcaption{Novec/water}
    \end{subfigure}
    \caption{Mean (\cref{eq:l1error}, dark color) and maximum
        (\cref{eq:maxerror}, light color) velocity errors over 
        time $t>\frac{1}{2}t_\text{max}$ from the 2D stationary droplet summerized as violin plots. For each
        fluid pairing, the results of different solvers and resolutions are displayed.}
    \label{fig:violinPlots_2DstationaryD}
\end{figure}

%% file: figures/latex-included/stationaryDrop3D_violinPlots.tex
\begin{figure}
    \centering
     \begin{subfigure}[b]{\textwidth}
        \includegraphics[width=\textwidth]{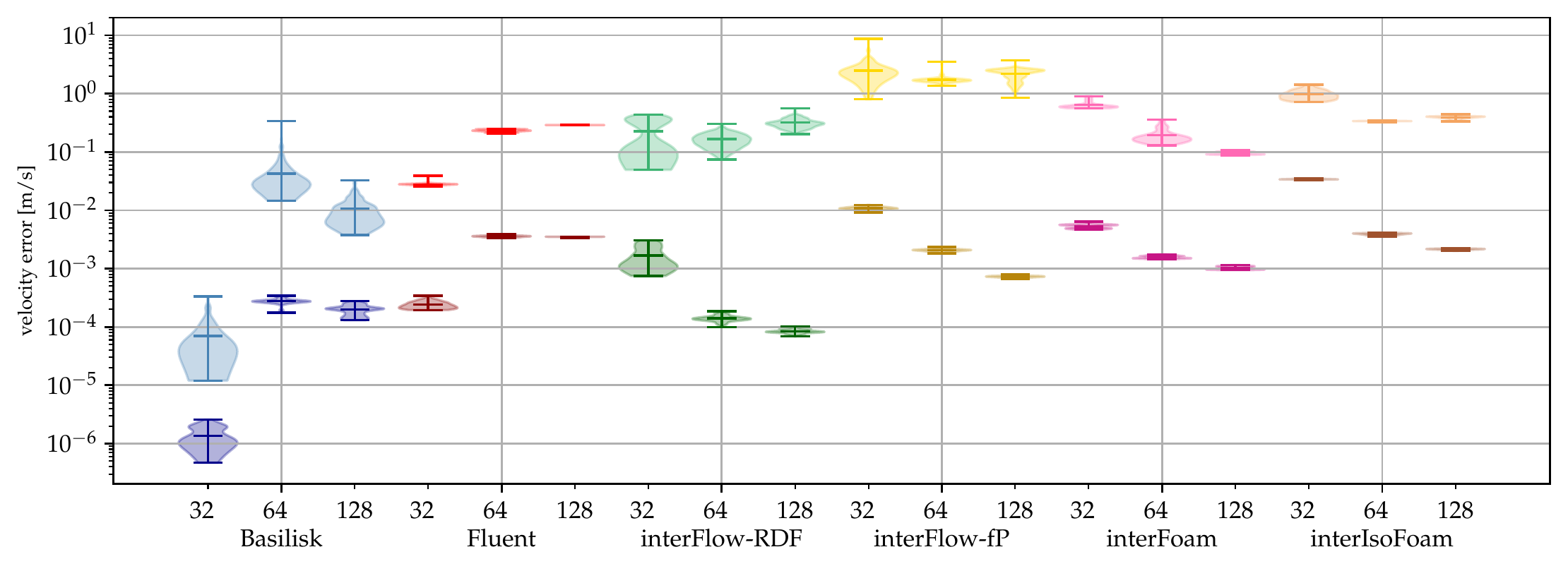}    
        \subcaption{water/air}
    \end{subfigure}
     \begin{subfigure}[b]{\textwidth}
        \includegraphics[width=\textwidth]{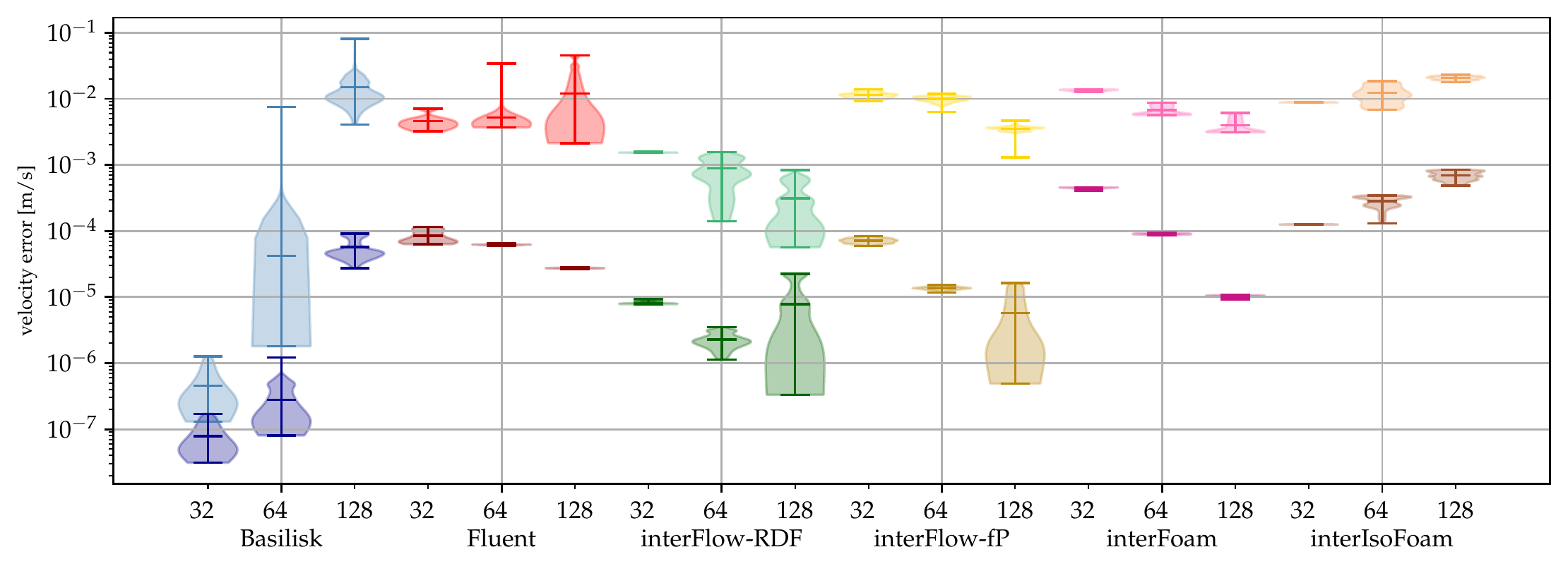}
        \subcaption{Ravenol/air}        
    \end{subfigure}
     \begin{subfigure}[b]{\textwidth}
        \includegraphics[width=\textwidth]{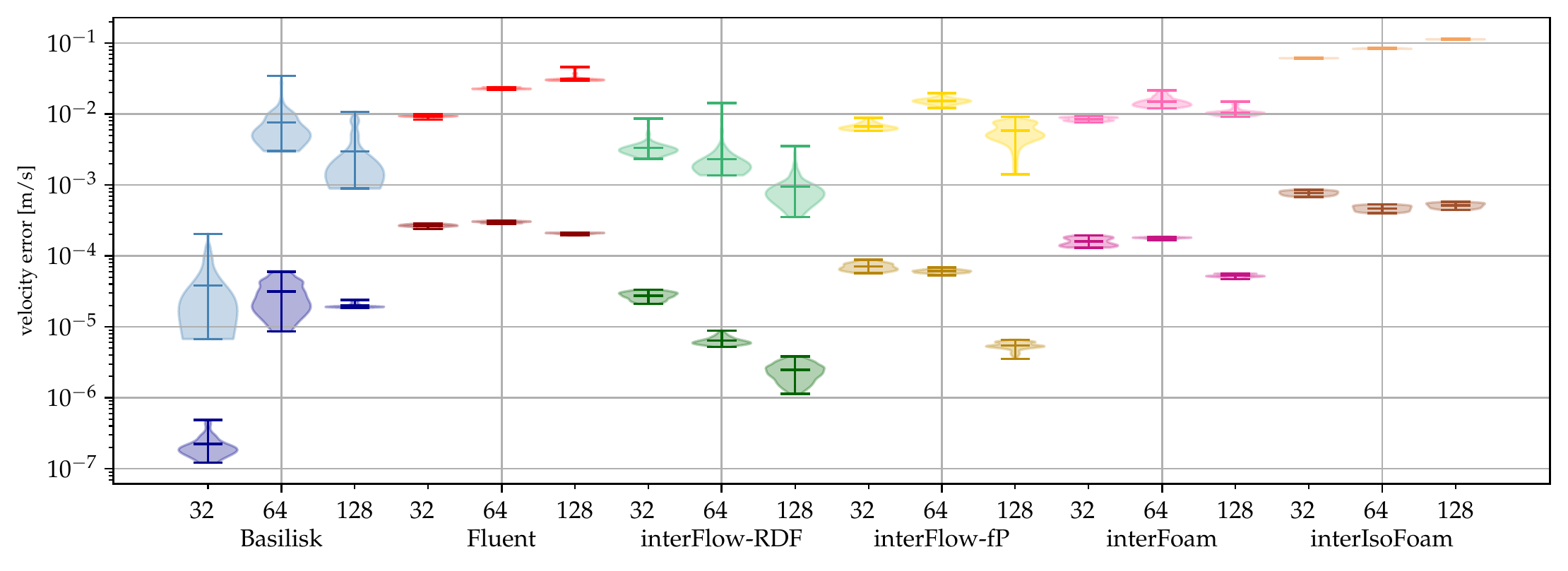}
        \subcaption{Novec/water}
    \end{subfigure}
    \caption{Mean (\cref{eq:l1error}, dark color) and maximum
        (\cref{eq:maxerror}, light color) velocity errors over 
        time $t>\frac{1}{2}t_\text{max}$ from the 3D stationary droplet summerized as violin plots. For each
        fluid pairing, the results of different solvers and resolutions are displayed.}
    \label{fig:violinPlots_3DstationaryD}
\end{figure}

%% file: figures/latex-included/translatingDrop2D_violinPlots.tex
\begin{figure}
    \centering
     \begin{subfigure}[b]{\textwidth}
        \includegraphics[width=\textwidth]{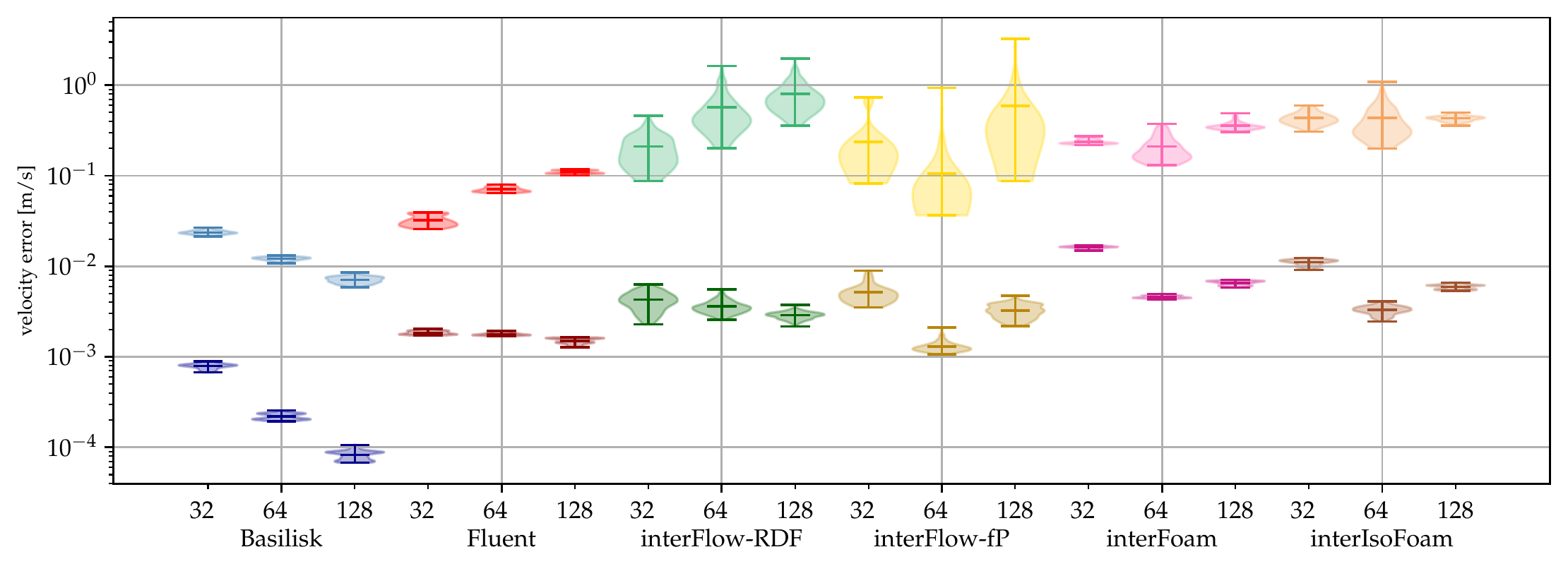}    
        \subcaption{water/air}
    \end{subfigure}
     \begin{subfigure}[b]{\textwidth}
        \includegraphics[width=\textwidth]{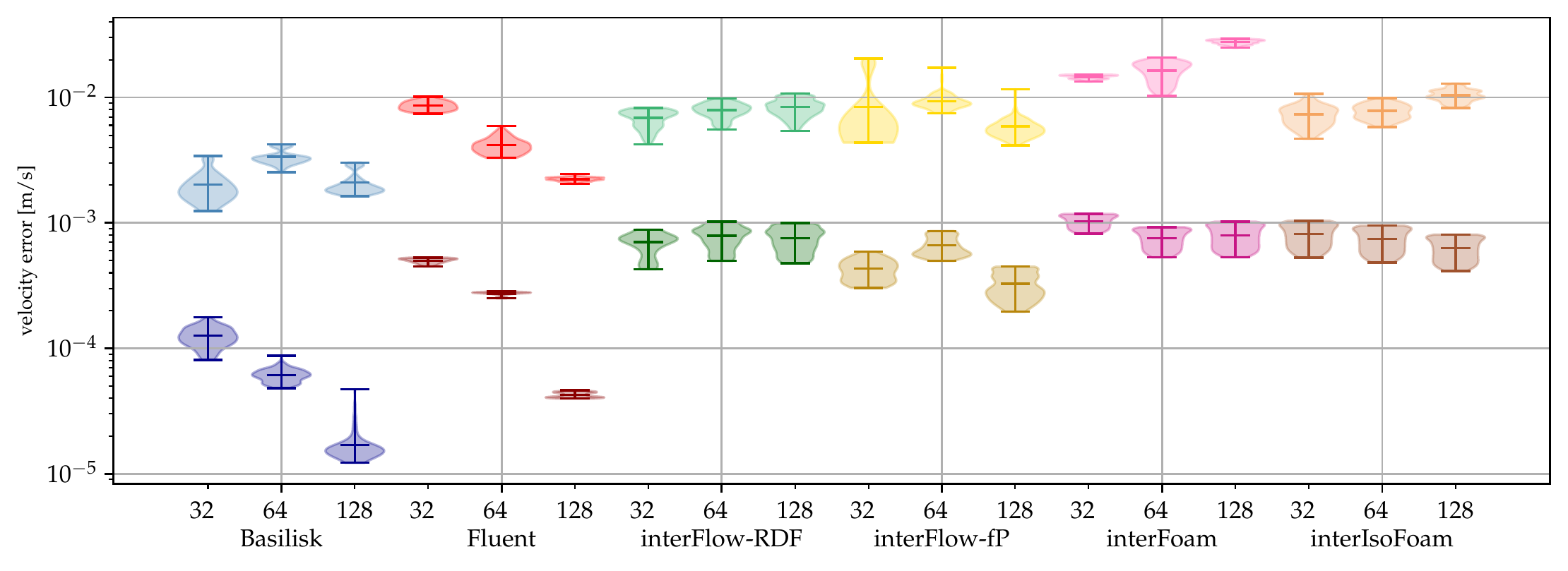}
        \subcaption{Ravenol/air}        
    \end{subfigure}
     \begin{subfigure}[b]{\textwidth}
        \includegraphics[width=\textwidth]{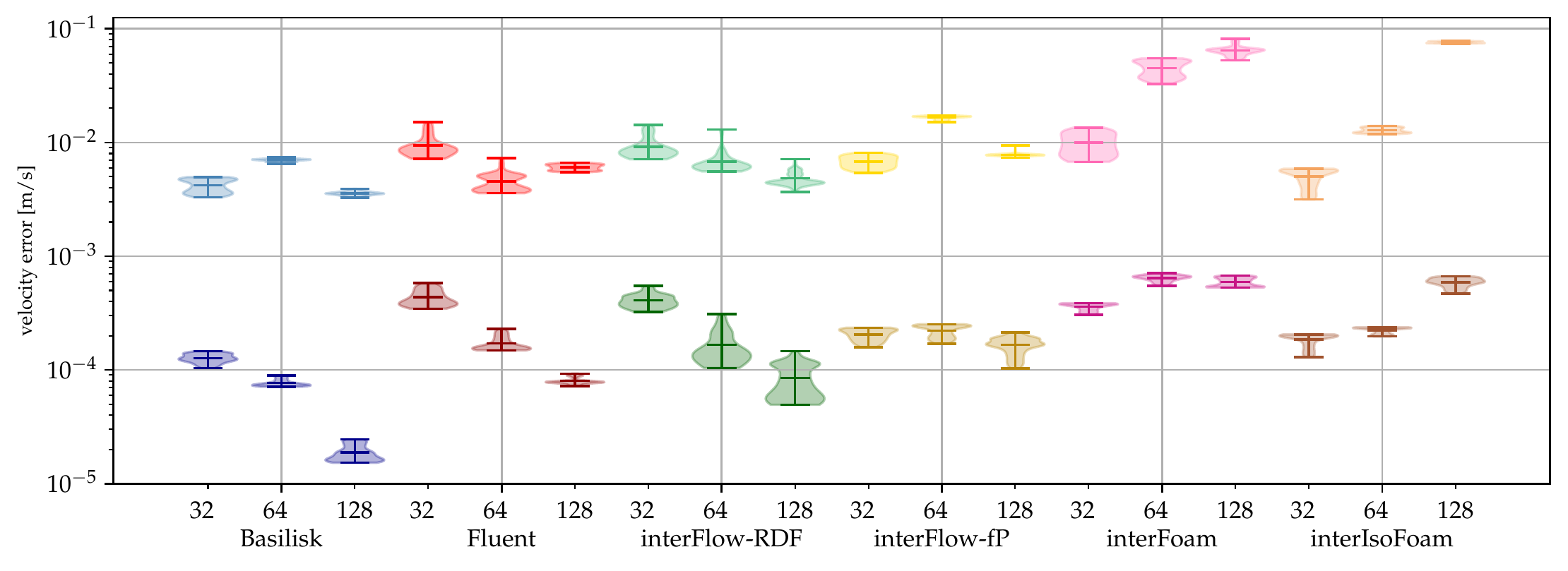}
        \subcaption{Novec/water}
    \end{subfigure}
    \caption{Mean (\cref{eq:l1error}, dark color) and maximum
        (\cref{eq:maxerror}, light color) velocity errors over 
        time $t>\frac{1}{2}t_\text{max}$ from the 2D translating droplet summerized as violin plots. For each
        fluid pairing, the results of different solvers and resolutions are displayed.}
    \label{fig:violinPlots_2DtranslatingD}
\end{figure}

%% file: figures/latex-included/translatingDrop3D_violinPlots.tex
\begin{figure}
    \centering
     \begin{subfigure}[b]{\textwidth}
        \includegraphics[width=\textwidth]{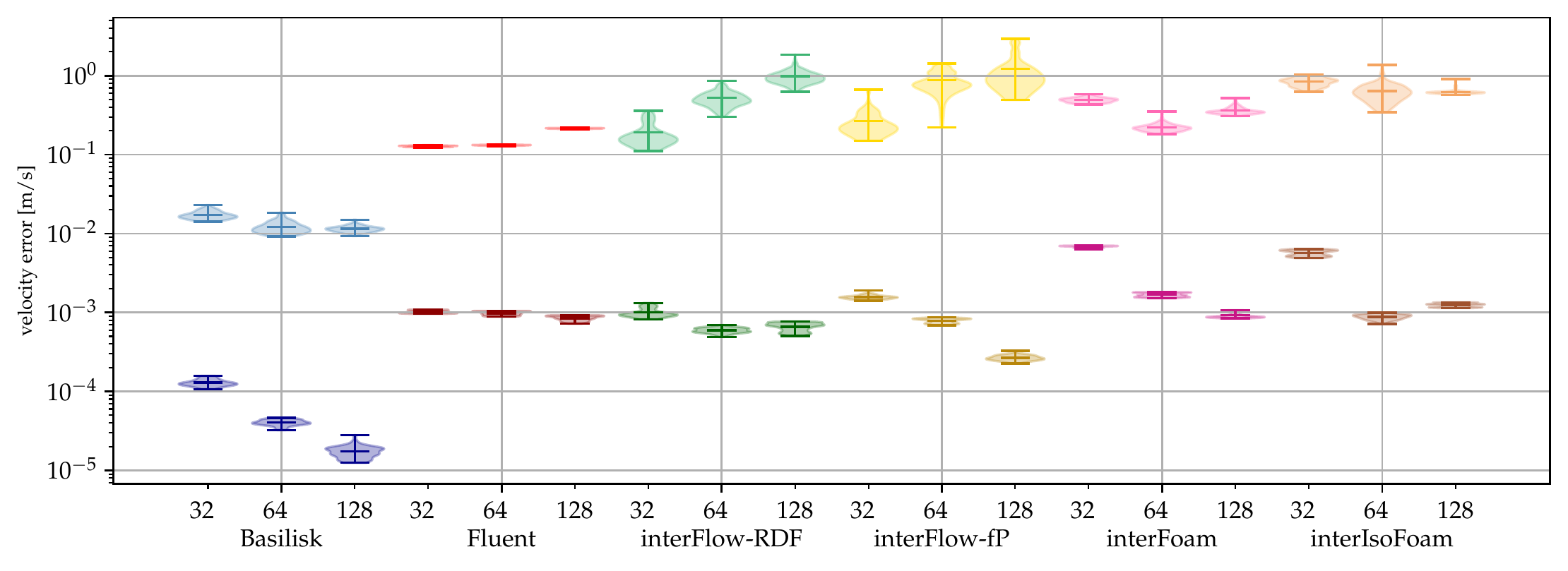}    
        \subcaption{water/air}
    \end{subfigure}
     \begin{subfigure}[b]{\textwidth}
        \includegraphics[width=\textwidth]{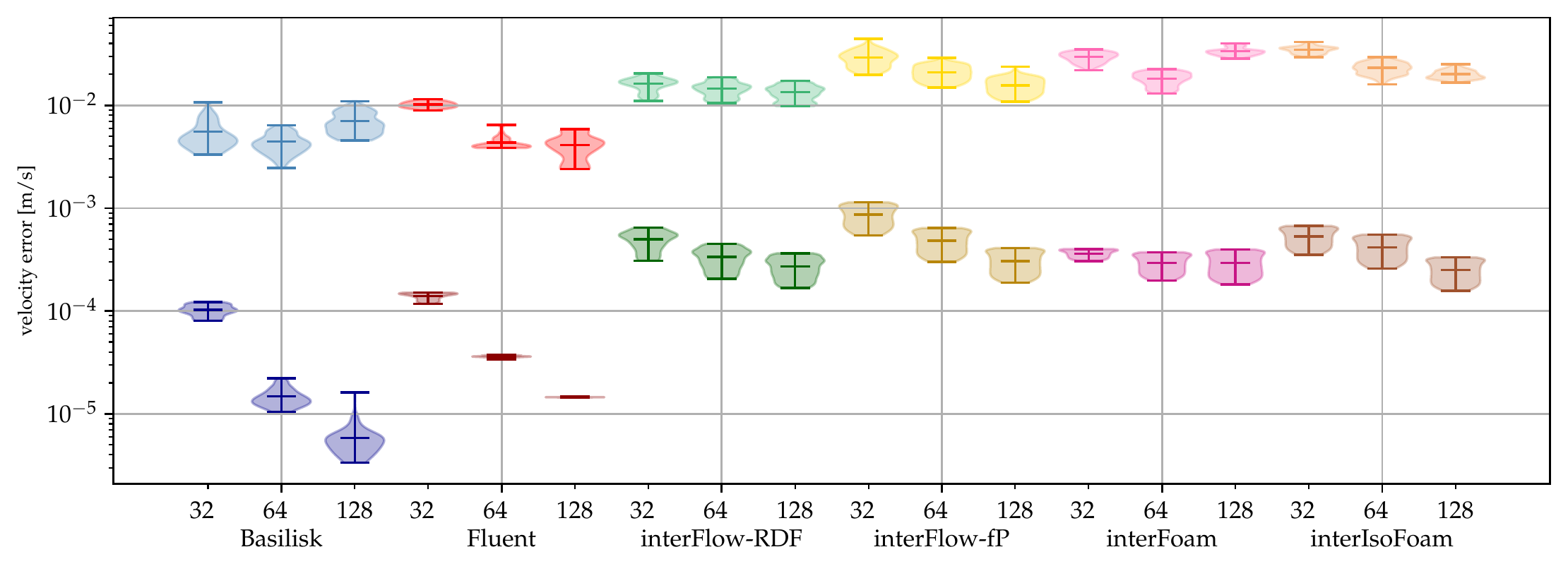}
        \subcaption{Ravenol/air}        
    \end{subfigure}
     \begin{subfigure}[b]{\textwidth}
        \includegraphics[width=\textwidth]{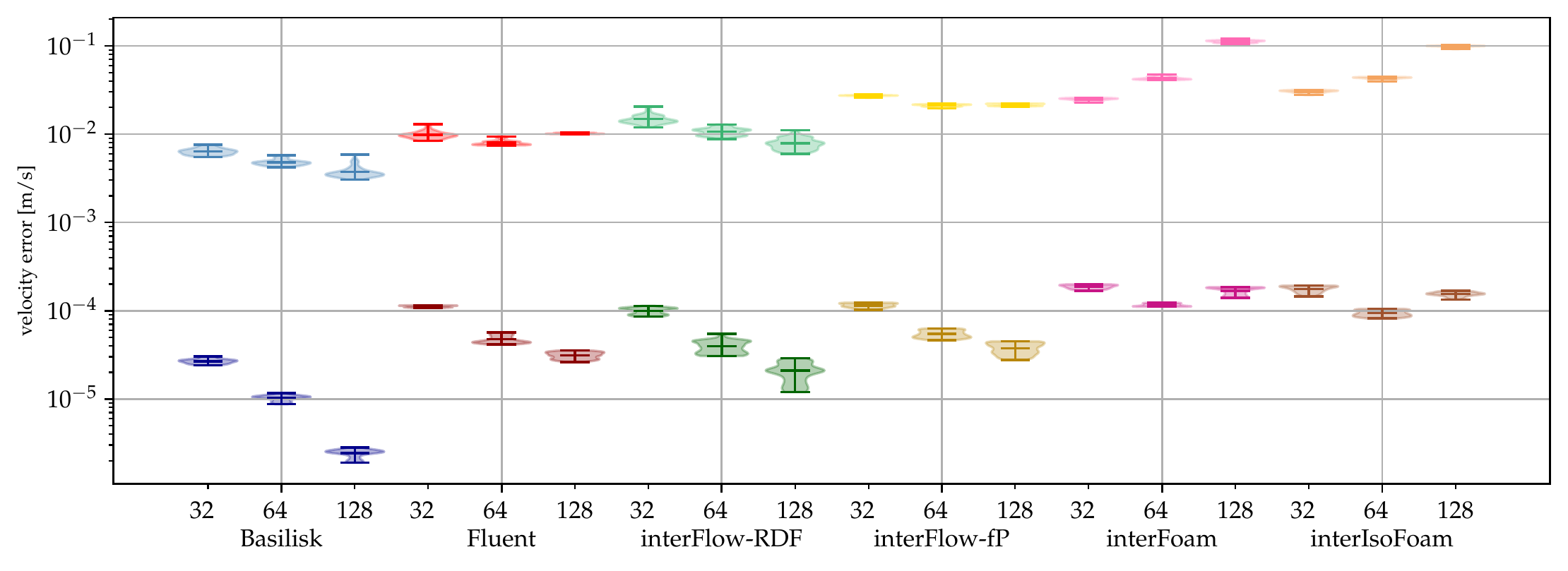}
        \subcaption{Novec/water}
    \end{subfigure}
    \caption{Mean (\cref{eq:l1error}, dark color) and maximum
        (\cref{eq:maxerror}, light color) velocity errors over 
        time $t>t_\text{max}$ from the 3D translating droplet summerized as violin plots. For each
        fluid pairing, the results of different solvers and resolutions are displayed.}
    \label{fig:violinPlots_3DtranslatingD}
\end{figure}

%% file: figures/latex-included/capillaryWave2D.tex
\begin{figure}[htb]
\centering
     \begin{subfigure}[b]{0.495\textwidth}
         \centering
         \includegraphics[width=\textwidth]{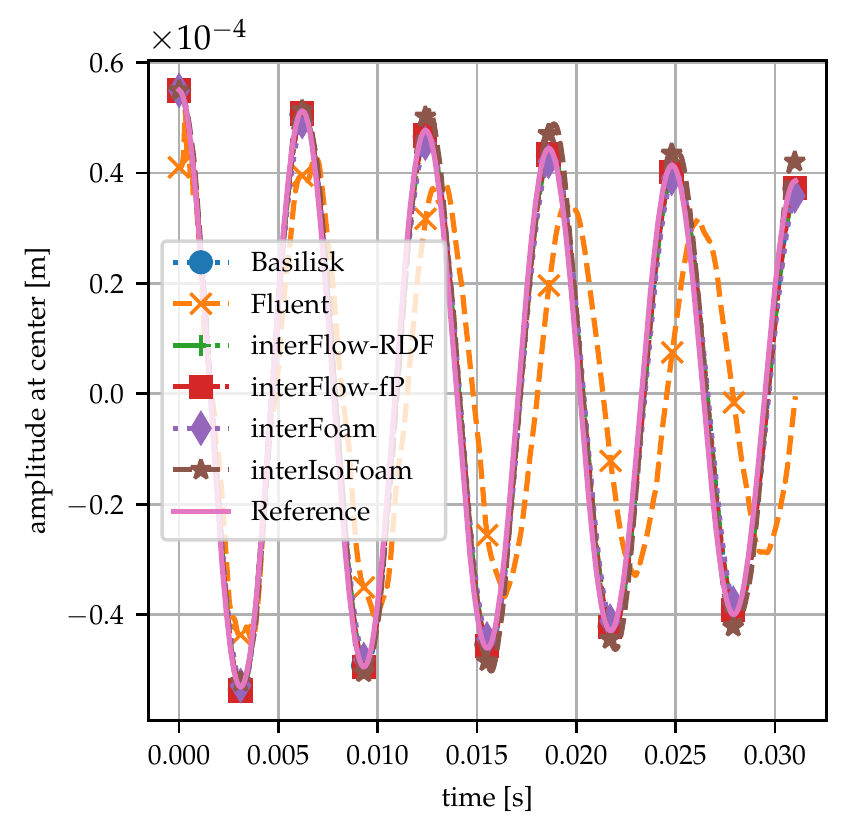}
         \caption{water/air, res: 128}
         \label{fig:oscillatingWave2D_oscillating_wave_2D_fluid_pairing__water-air_resolution__128_}
     \end{subfigure}
     \hfill
     \begin{subfigure}[b]{0.495\textwidth}
         \centering
         \includegraphics[width=\textwidth]{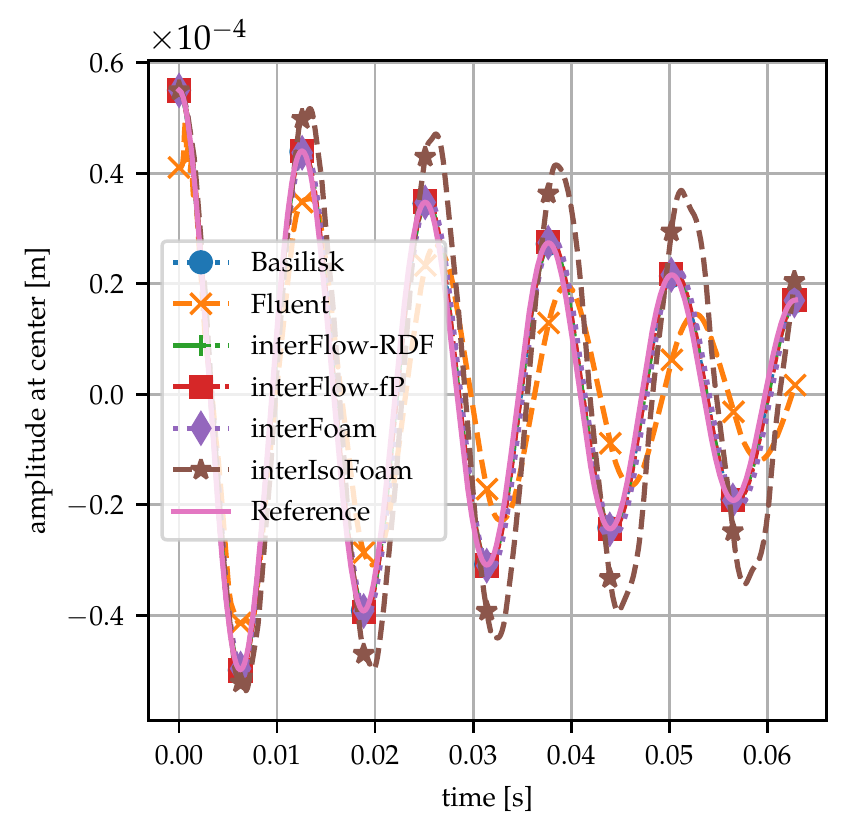}
         \caption{Novec/water, res: 128}
         \label{fig:oscillatingWave2D_oscillating_wave_2D_fluid_pairing__oil_novec7500-water_resolution__128_}
     \end{subfigure}     
    \caption{Tracking the film height at center over time for oscillating wave. The reference is given by the analytical solution by~\cite{Prosperetti1981}.}
    \label{fig:temporal_evolution_2DoscWave}
\end{figure}

%% file: figures/latex-included/oscillatingWave2D_accuPlots.tex
\begin{figure}[tb]
    \centering
     \begin{subfigure}[b]{0.49\textwidth}
        \includegraphics[width=\textwidth]{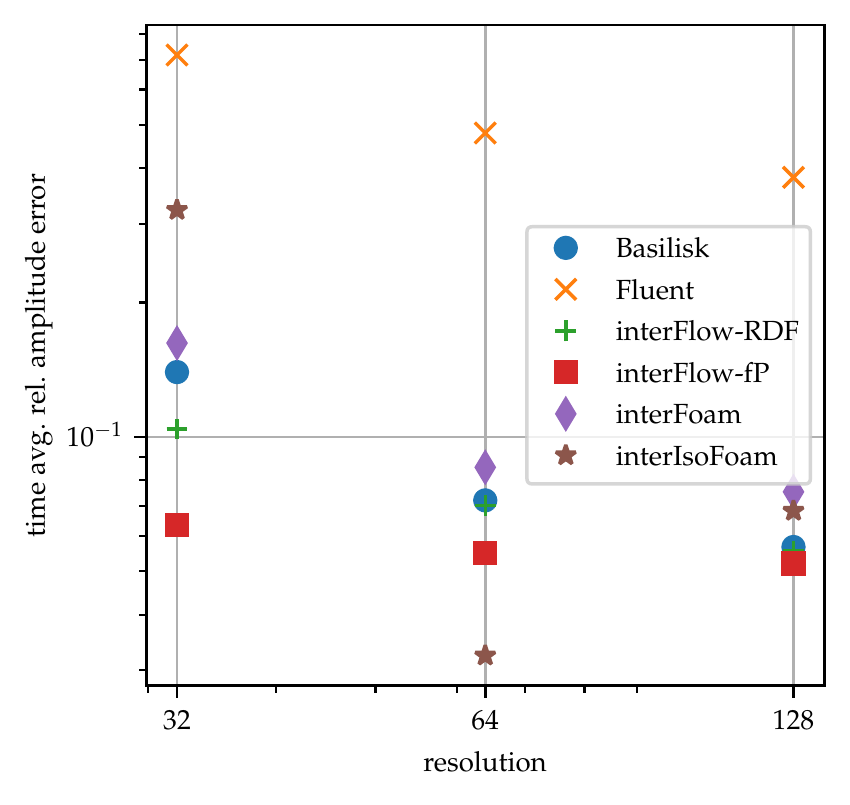}    
        \subcaption{water/air}
    \end{subfigure}
     \begin{subfigure}[b]{0.49\textwidth}
        \includegraphics[width=\textwidth]{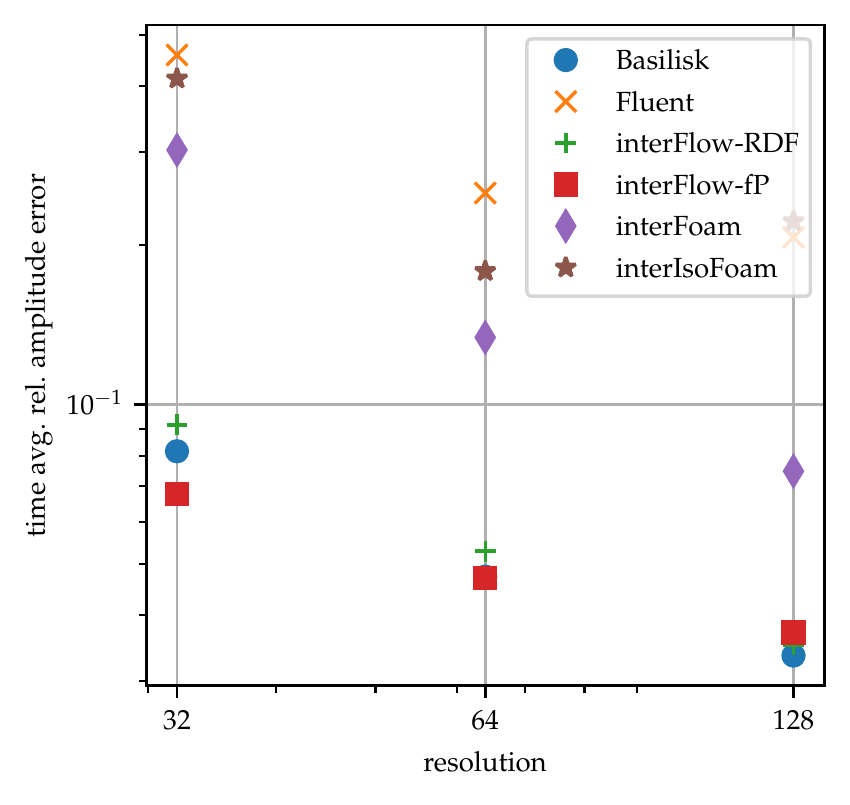} 
        \subcaption{Novec/water}
    \end{subfigure}
    \caption{Time averaged relative difference between simulated wave height and analytical solution for a 2D oscillating wave.}
    \label{fig:time_averaged_2DoscWave}
\end{figure}

%% file: figures/latex-included/oscillatingDrop3D.tex
\begin{figure}[htb]
     \begin{subfigure}[b]{0.495\textwidth}
         \centering
         \includegraphics[width=\textwidth]{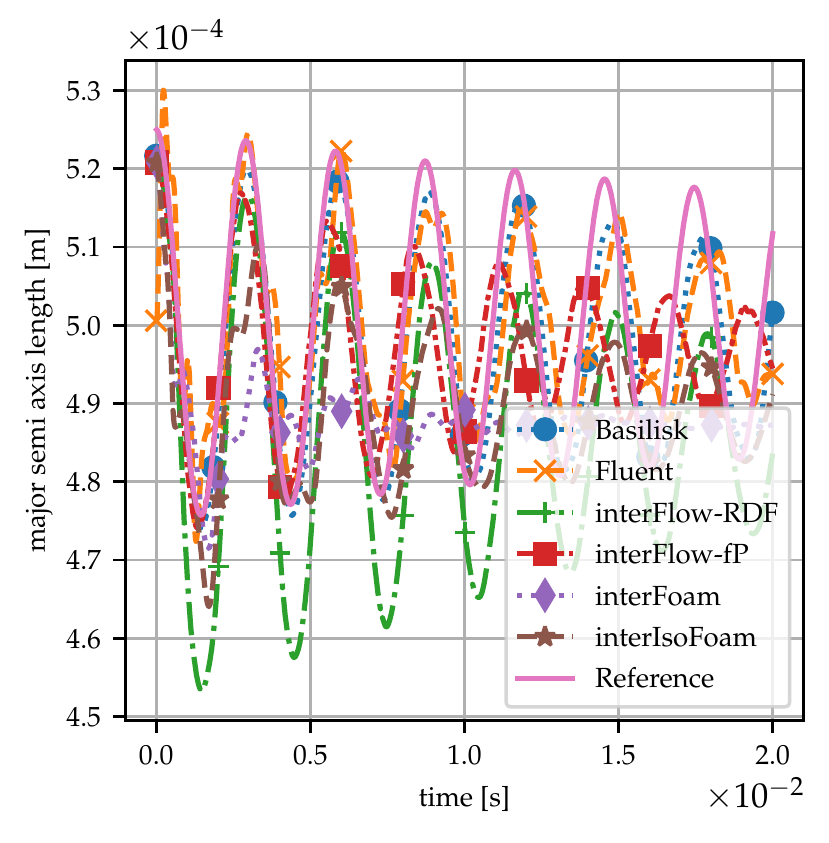}
         \caption{Water/air, res: 100}
         \label{fig:oscillatingDroplet3D_oscillating_droplet_3D_fluid_pairing__water-air_resolution__100_}
     \end{subfigure}
     \hfill
     \begin{subfigure}[b]{0.495\textwidth}
         \centering
         \includegraphics[width=\textwidth]{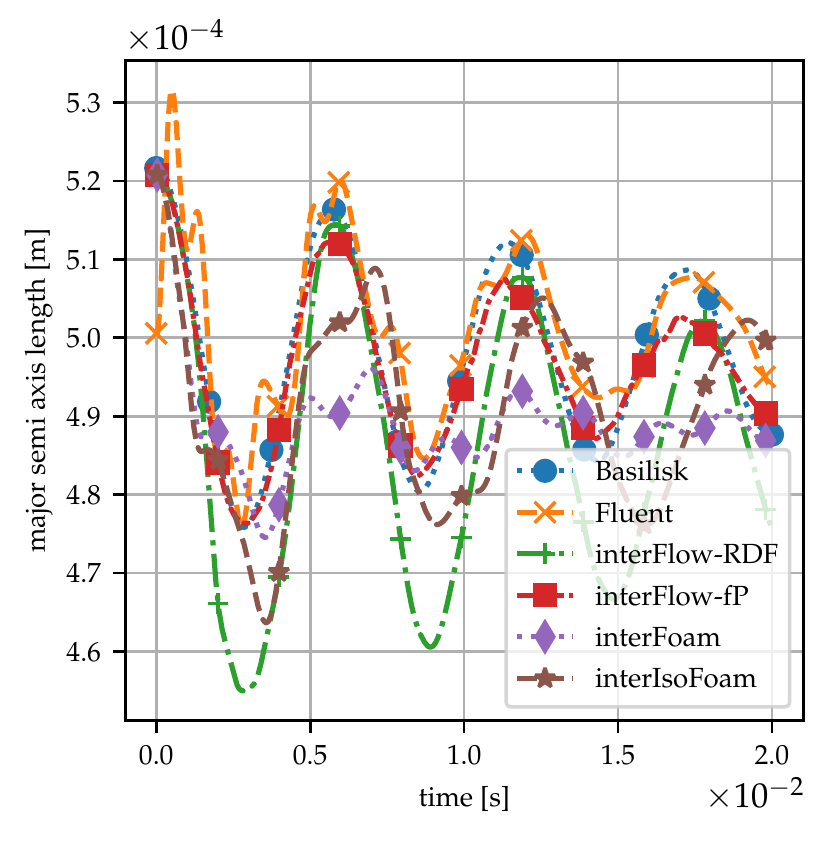}
         \caption{Novec/water, res: 100}
         \label{fig:oscillatingDroplet3D_oscillating_droplet_3D_fluid_pairing__oil_novec7500-water_resolution__100_}
     \end{subfigure}     

    \caption{Tracking the initially longest semi-axis in 3D over time for the oscillating droplet.}
    \label{fig:temporal_evolution_oscDrop3D}
\end{figure}

%% file: figures/latex-included/oscillatingDrop3D_accuPlots.tex
\begin{figure}
    \centering

    \includegraphics[width=0.6\textwidth]{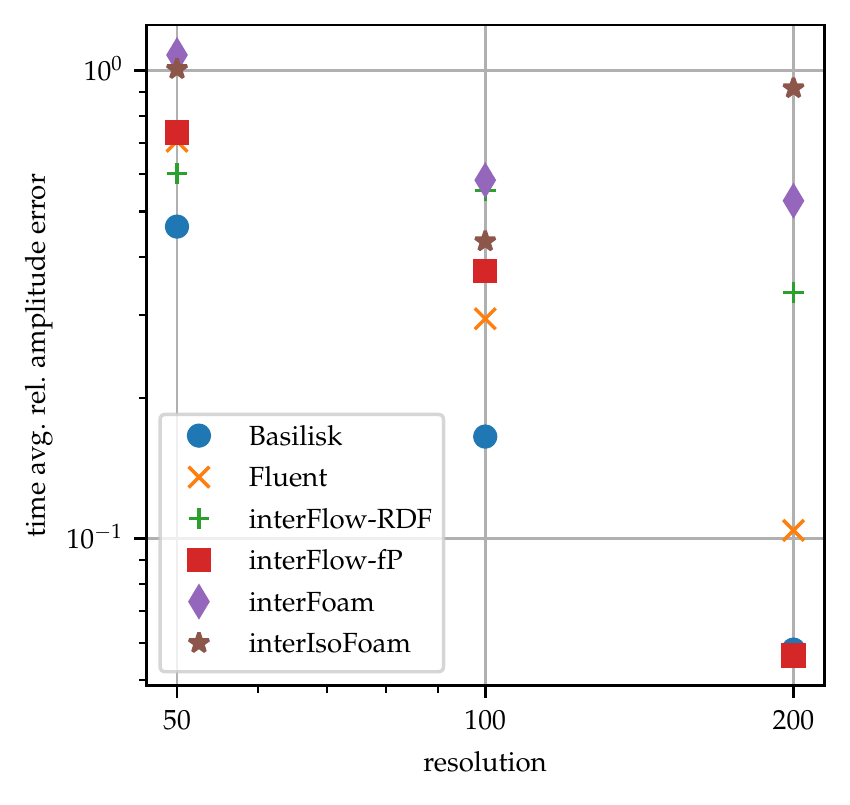}    

    \caption{Time averaged relative difference between simulated initially longest semi-axis and the analytical solution for 3D oscillating drop for water/air.}
    \label{fig:accuPlots_3DoscD}
\end{figure}

%% file: sections/conclusions.tex
\section{Conclusions}
\label{sec:concl}

We provide a basis for continuously benchmarking numerical methods for surface-tension-driven incompressible two-phase flows that discretize the single-field formulation of two-phase Navier-Stokes equations. The post-processing implemented in Jupyter notebooks can be re-used without modification by methods developed in other software to compare with our benchmark results directly. Providing the benchmark suite as a reusable open-source project can save significant research time.

Overall, the benchmark results demonstrate a need for significant improvements in the approximation of surface tension forces, especially for the unstructured VoF methods available in OpenFOAM and Fluent. The structured VoF method in Basilisk delivers the best results with the height-function curvature approximation, but not for all fluid pairings, and with higher accuracy in 2D compared to 3D. 

This study also highlights the importance of an accurate field initialization to enable an acceptable solver performance.

In future work, we will re-use the benchmark suite to continuously compare new approximation methods for surface tension forces with other methods. We invite the multiphase community to participate in this direct comparison, following the instructions in the linked code repository \cite{BoschGitHub}.






\section{Acknowledgments}

The last author acknowledges the funding received from the German Research Foundation (DFG)- Project-ID 265191195 - SFB 1194.

%% file: ms.bbl
\begin{thebibliography}{44}
\expandafter\ifx\csname natexlab\endcsname\relax\def\natexlab#1{#1}\fi
\providecommand{\url}[1]{\texttt{#1}}
\providecommand{\href}[2]{#2}
\providecommand{\path}[1]{#1}
\providecommand{\DOIprefix}{doi:}
\providecommand{\ArXivprefix}{arXiv:}
\providecommand{\URLprefix}{URL: }
\providecommand{\Pubmedprefix}{pmid:}
\providecommand{\doi}[1]{\href{http://dx.doi.org/#1}{\path{#1}}}
\providecommand{\Pubmed}[1]{\href{pmid:#1}{\path{#1}}}
\providecommand{\bibinfo}[2]{#2}
\ifx\xfnm\relax \def\xfnm[#1]{\unskip,\space#1}\fi
\bibitem[{Hirt and Nichols(1981)}]{hirt1981volume}
\bibinfo{author}{C.~W. Hirt}, \bibinfo{author}{B.~D. Nichols},
\newblock \bibinfo{title}{Volume of fluid (vof) method for the dynamics of free
  boundaries},
\newblock \bibinfo{journal}{Journal of Computational Physics}
  \bibinfo{volume}{39} (\bibinfo{year}{1981}) \bibinfo{pages}{201--225}.
\bibitem[{Mari{\'c} et~al.(2020)Mari{\'c}, Kothe, and
  Bothe}]{maric2020unstructured}
\bibinfo{author}{T.~Mari{\'c}}, \bibinfo{author}{D.~B. Kothe},
  \bibinfo{author}{D.~Bothe},
\newblock \bibinfo{title}{Unstructured un-split geometrical volume-of-fluid
  methods--a review},
\newblock \bibinfo{journal}{Journal of Computational Physics}
  \bibinfo{volume}{420} (\bibinfo{year}{2020}) \bibinfo{pages}{109695}.
\bibitem[{Popinet(2018)}]{popinet2018numerical}
\bibinfo{author}{S.~Popinet},
\newblock \bibinfo{title}{Numerical models of surface tension},
\newblock \bibinfo{journal}{Annual Review of Fluid Mechanics}
  \bibinfo{volume}{50} (\bibinfo{year}{2018}) \bibinfo{pages}{49--75}.
\bibitem[{Oberkampf and Roy(2010)}]{Oberkampf2010verification}
\bibinfo{author}{W.~L. Oberkampf}, \bibinfo{author}{C.~J. Roy},
  \bibinfo{title}{Verification and validation in scientific computing},
  \bibinfo{publisher}{Cambridge University Press}, \bibinfo{year}{2010}.
\bibitem[{Grambow et~al.(2019)Grambow, Lehmann, and Bermbach}]{Grambow2019}
\bibinfo{author}{M.~Grambow}, \bibinfo{author}{F.~Lehmann},
  \bibinfo{author}{D.~Bermbach},
\newblock \bibinfo{title}{Continuous {Benchmarking}: {Using} {System}
  {Benchmarking} in {Build} {Pipelines}},
\newblock in: \bibinfo{booktitle}{2019 {IEEE} {International} {Conference} on
  {Cloud} {Engineering} ({IC2E})}, \bibinfo{publisher}{IEEE},
  \bibinfo{address}{Prague, Czech Republic}, \bibinfo{year}{2019}, pp.
  \bibinfo{pages}{241--246}. \URLprefix
  \url{https://ieeexplore.ieee.org/document/8790186/}.
  \DOIprefix\doi{10.1109/IC2E.2019.00039}.
\bibitem[{{OpenCFD Ldt.}(2022)}]{OpenFOAMcode}
\bibinfo{author}{{OpenCFD Ldt.}}, \bibinfo{title}{{OpenFOAM Repository}},
  \bibinfo{howpublished}{\url{https://develop.openfoam.com/Development/openfoam/-/tree/OpenFOAM-v2112}},
  \bibinfo{year}{2022}. \bibinfo{note}{Git tag OpenFOAM-v2112}.
\bibitem[{Scheufler and Roenby(2021)}]{scheufler2021twophaseflow}
\bibinfo{author}{H.~Scheufler}, \bibinfo{author}{J.~Roenby},
\newblock \bibinfo{title}{{TwoPhaseFlow: An OpenFOAM based framework for
  development of two phase flow solvers}},
\newblock \bibinfo{journal}{arXiv preprint arXiv:2103.00870}
  (\bibinfo{year}{2021}).
\bibitem[{Two(2022)}]{TwoPhaseFlowCode}
\bibinfo{title}{{TwoPhaseFlow Repository, version of2112}},
  \bibinfo{howpublished}{\url{https://github.com/DLR-RY/TwoPhaseFlow/tree/of2112}},
  \bibinfo{year}{2022}. \bibinfo{note}{Git commit:
  cbaf314577987db34a9e41bb4bdca511c40d080}.
\bibitem[{{{ANSYS, Inc.}}(2020)}]{ansysfluentuser2020r1}
\bibinfo{author}{{{ANSYS, Inc.}}}, \bibinfo{title}{{{Ansys Fluent User Guide
  2020 R1}}}, \bibinfo{year}{2020}.
\bibitem[{{S. Popinet and collaborators}(2015)}]{basilisk}
\bibinfo{author}{{S. Popinet and collaborators}}, \bibinfo{title}{Basilisk},
  \bibinfo{howpublished}{\url{http://basilisk.fr}}, \bibinfo{year}{since 2015}.
  \bibinfo{note}{Accessed: 2022-10-13}.
\bibitem[{Lippert et~al.(1 28)Lippert, Tolle, Dörr, and Maric}]{BenchmarkData}
\bibinfo{author}{A.~Lippert}, \bibinfo{author}{T.~Tolle},
  \bibinfo{author}{A.~Dörr}, \bibinfo{author}{T.~Maric}, \bibinfo{title}{A
  benchmark for surface-tension-driven incompressible two-phase flows -
  benchmark data}, \bibinfo{year}{2022-11-28}. \URLprefix
  \url{https://tudatalib.ulb.tu-darmstadt.de/handle/tudatalib/3630}.
  \DOIprefix\doi{10.48328/tudatalib-989}.
\bibitem[{{{Robert Bosch GmbH}}(2022)}]{BoschGitHub}
\bibinfo{author}{{{Robert Bosch GmbH}}},
  \bibinfo{title}{{BoschResearch/sepMutliphaseFoam, git
  tree/publications/ST-VoF-benchmark}},
  \bibinfo{howpublished}{\url{https://github.com/boschresearch/sepMultiphaseFoam/tree/publications/ST-VoF-benchmark}},
  \bibinfo{year}{2022}. \bibinfo{note}{Created: 2022-11-20}.
\bibitem[{Tryggvason et~al.(2011)Tryggvason, Scardovelli, and
  Zaleski}]{Tryggvason2011}
\bibinfo{author}{G.~Tryggvason}, \bibinfo{author}{R.~Scardovelli},
  \bibinfo{author}{S.~Zaleski}, \bibinfo{title}{Direct numerical simulations of
  gas--liquid multiphase flows}, \bibinfo{publisher}{Cambridge university
  press}, \bibinfo{year}{2011}.
\bibitem[{Jasak(1996)}]{Jasak1996}
\bibinfo{author}{H.~Jasak}, \bibinfo{title}{Error analysis and estimation for
  the finite volume method with applications to fluid flows.}, Ph.D. thesis,
  Imperial College London (University of London), \bibinfo{year}{1996}.
\bibitem[{Hirsch(2007)}]{Hirsch2007}
\bibinfo{author}{C.~Hirsch}, \bibinfo{title}{Numerical computation of internal
  and external flows: The fundamentals of computational fluid dynamics},
  \bibinfo{publisher}{Elsevier}, \bibinfo{year}{2007}.
\bibitem[{Maric et~al.(2014)Maric, Hopken, and Mooney}]{Maric2014}
\bibinfo{author}{T.~Maric}, \bibinfo{author}{J.~Hopken},
  \bibinfo{author}{K.~Mooney}, \bibinfo{title}{The OpenFOAM technology primer},
  \bibinfo{publisher}{Sourceflux}, \bibinfo{year}{2014}.
\bibitem[{Moukalled et~al.(2016)Moukalled, Mangani, and
  Darwish}]{Moukalled2016}
\bibinfo{author}{F.~Moukalled}, \bibinfo{author}{L.~Mangani},
  \bibinfo{author}{M.~Darwish}, \bibinfo{title}{The finite volume method},
  \bibinfo{publisher}{Springer}, \bibinfo{year}{2016}.
\bibitem[{Denner et~al.(2022)Denner, Evrard, and van
  Wachem}]{Denner2022implicit}
\bibinfo{author}{F.~Denner}, \bibinfo{author}{F.~Evrard},
  \bibinfo{author}{B.~van Wachem},
\newblock \bibinfo{title}{Breaching the capillary time-step constraint using a
  coupled vof method with implicit surface tension},
\newblock \bibinfo{journal}{Journal of Computational Physics}
  \bibinfo{volume}{459} (\bibinfo{year}{2022}) \bibinfo{pages}{111128}.
\bibitem[{Deshpande et~al.(2012)Deshpande, Anumolu, and
  Trujillo}]{Deshphande2012}
\bibinfo{author}{S.~Deshpande}, \bibinfo{author}{L.~Anumolu},
  \bibinfo{author}{M.~Trujillo},
\newblock \bibinfo{title}{Evaluating the performance of the two-phase flow
  solver interfoam},
\newblock \bibinfo{journal}{Computational Science \& Discovery}
  \bibinfo{volume}{5} (\bibinfo{year}{2012}).
  \DOIprefix\doi{10.1088/1749-4699/5/1/014016}.
\bibitem[{Chen et~al.(2022)Chen, Xie, and Xiao}]{Chen2022}
\bibinfo{author}{D.~Chen}, \bibinfo{author}{B.~Xie}, \bibinfo{author}{F.~Xiao},
\newblock \bibinfo{title}{Revisit to the thinc/qq scheme: Recent progress to
  improve accuracy and robustness},
\newblock \bibinfo{journal}{International Journal for Numerical Methods in
  Fluids} \bibinfo{volume}{94} (\bibinfo{year}{2022})
  \bibinfo{pages}{719--755}.
\bibitem[{Roenby et~al.(2016)Roenby, Bredmose, and Jasak}]{Roenby2016}
\bibinfo{author}{J.~Roenby}, \bibinfo{author}{H.~Bredmose},
  \bibinfo{author}{H.~Jasak},
\newblock \bibinfo{title}{A {Computational} {Method} for {Sharp} {Interface}
  {Advection}},
\newblock \bibinfo{journal}{Royal Society Open Science} \bibinfo{volume}{3}
  (\bibinfo{year}{2016}). \URLprefix \url{http://arxiv.org/abs/1601.05392}.
  \DOIprefix\doi{10.1098/rsos.160405}.
\bibitem[{Scheufler and Roenby(2019)}]{Scheufler2019}
\bibinfo{author}{H.~Scheufler}, \bibinfo{author}{J.~Roenby},
\newblock \bibinfo{title}{Accurate and efficient surface reconstruction from
  volume fraction data on general meshes},
\newblock \bibinfo{journal}{Journal of Computational Physics}
  \bibinfo{volume}{383} (\bibinfo{year}{2019}) \bibinfo{pages}{1--23}.
  \URLprefix
  \url{https://www.sciencedirect.com/science/article/pii/S0021999119300269}.
  \DOIprefix\doi{10.1016/j.jcp.2019.01.009}, \bibinfo{note}{arXiv: 1801.05382
  Publisher: Academic Press}.
\bibitem[{Popinet(2009)}]{popinet2009}
\bibinfo{author}{S.~Popinet},
\newblock \bibinfo{title}{An accurate adaptive solver for
  surface-tension-driven interfacial flows},
\newblock \bibinfo{journal}{Journal of Computational Physics}
  \bibinfo{volume}{228} (\bibinfo{year}{2009}) \bibinfo{pages}{5838--5866}.
  \DOIprefix\doi{doi.org/10.1016/j.jcp.2009.04.042}.
\bibitem[{{{ANSYS, Inc.}}(2020)}]{ansysfluenttheory2020r1}
\bibinfo{author}{{{ANSYS, Inc.}}}, \bibinfo{title}{{{Ansys Fluent Theory Guide
  2020 R1}}}, \bibinfo{year}{2020}.
\bibitem[{Brackbill et~al.(1992)Brackbill, Kothe, and
  Zemach}]{brackbill1992continuum}
\bibinfo{author}{J.~U. Brackbill}, \bibinfo{author}{D.~B. Kothe},
  \bibinfo{author}{C.~Zemach},
\newblock \bibinfo{title}{A continuum method for modeling surface tension},
\newblock \bibinfo{journal}{Journal of computational physics}
  \bibinfo{volume}{100} (\bibinfo{year}{1992}) \bibinfo{pages}{335--354}.
\bibitem[{Torrey et~al.(1985)Torrey, Cloutman, Mjolsness, and
  Hirt}]{torrey1985}
\bibinfo{author}{M.~D. Torrey}, \bibinfo{author}{L.~D. Cloutman},
  \bibinfo{author}{R.~C. Mjolsness}, \bibinfo{author}{C.~W. Hirt},
  \bibinfo{title}{NASA-VOF2D: a computer program for incompressible flows with
  free surfaces}, \bibinfo{type}{Technical Report}, Los Alamos National Lab.,
  NM (USA), \bibinfo{year}{1985}. \URLprefix
  \url{https://www.osti.gov/biblio/5934123}.
\bibitem[{Ivey and Moin(2015)}]{ivey2015}
\bibinfo{author}{C.~B. Ivey}, \bibinfo{author}{P.~Moin},
\newblock \bibinfo{title}{Accurate interface normal and curvature estimates on
  three-dimensional unstructured non-convex polyhedral meshes},
\newblock \bibinfo{journal}{Journal of Computational Physics}
  \bibinfo{volume}{300} (\bibinfo{year}{2015}) \bibinfo{pages}{365--386}.
  \DOIprefix\doi{doi.org/10.1016/j.jcp.2015.07.055}.
\bibitem[{Jibben et~al.(2019)Jibben, Carlson, and Francois}]{jibben2019}
\bibinfo{author}{Z.~Jibben}, \bibinfo{author}{N.~Carlson},
  \bibinfo{author}{M.~Francois},
\newblock \bibinfo{title}{A paraboloid fitting technique for calculating
  curvature from piecewise-linear interface reconstructions on 3d unstructured
  meshes},
\newblock \bibinfo{journal}{Computers \& Mathematics with Applications}
  \bibinfo{volume}{78} (\bibinfo{year}{2019}) \bibinfo{pages}{643--653}.
  \DOIprefix\doi{10.1016/j.camwa.2018.09.009}, \bibinfo{note}{proceedings of
  the Eight International Conference on Numerical Methods for Multi-Material
  Fluid Flows (MULTIMAT 2017)}.
\bibitem[{Cummins et~al.(2005)Cummins, Francois, and Kothe}]{cummins2005}
\bibinfo{author}{S.~J. Cummins}, \bibinfo{author}{M.~M. Francois},
  \bibinfo{author}{D.~B. Kothe},
\newblock \bibinfo{title}{Estimating curvature from volume fractions},
\newblock \bibinfo{journal}{Computers \& Structures} \bibinfo{volume}{83}
  (\bibinfo{year}{2005}) \bibinfo{pages}{425--434}.
  \DOIprefix\doi{10.1016/j.compstruc.2004.08.017}, \bibinfo{note}{frontier of
  Multi-Phase Flow Analysis and Fluid-Structure}.
\bibitem[{Jureti{\'c}(2005)}]{juretic2005phd}
\bibinfo{author}{F.~Jureti{\'c}}, \bibinfo{title}{Error analysis in finite
  volume CFD}, Ph.D. thesis, Imperial College London (University of London),
  \bibinfo{year}{2005}.
\bibitem[{{{ANSYS, Inc.}}(2020)}]{ansysmeshinguser2020r1}
\bibinfo{author}{{{ANSYS, Inc.}}}, \bibinfo{title}{{{Ansys Meshing User Guide
  2020 R1}}}, \bibinfo{year}{2020}.
\bibitem[{{Christophe Geuzaine and Jean-Francois Remacle}(2020)}]{gmsh2020}
\bibinfo{author}{{Christophe Geuzaine and Jean-Francois Remacle}},
  \bibinfo{title}{Gmsh}, \bibinfo{year}{2020}. \URLprefix
  \url{http://http://gmsh.info/}.
\bibitem[{Riegel et~al.(2022)Riegel, Mayer, and van Havre}]{freecad2022}
\bibinfo{author}{J.~Riegel}, \bibinfo{author}{W.~Mayer},
  \bibinfo{author}{Y.~van Havre}, \bibinfo{title}{{FreeCAD (Version 0.19)}},
  \bibinfo{year}{2001-2022}. \URLprefix \url{http://www.freecadweb.org}.
\bibitem[{Lemmon et~al.(2022)Lemmon, Bell, Huber, and McLinden}]{Lemmon2022}
\bibinfo{author}{E.~W. Lemmon}, \bibinfo{author}{I.~H. Bell},
  \bibinfo{author}{M.~L. Huber}, \bibinfo{author}{M.~O. McLinden},
  \bibinfo{title}{NIST Chemistry WebBook, NIST Standard Reference Database
  Number 69}, \bibinfo{publisher}{National Institute of Standards and
  Technology, Gaithersburg MD}, \bibinfo{year}{2022}, p.
  \bibinfo{pages}{20899}. \URLprefix \url{https://webbook.nist.gov}.
\bibitem[{{VDI Gesellschaft}(2010)}]{VDI2010}
\bibinfo{editor}{{VDI Gesellschaft}} (Ed.), \bibinfo{title}{VDI Heat Atlas},
  VDI-Buch, \bibinfo{publisher}{Springer Berlin Heidelberg},
  \bibinfo{year}{2010}. \URLprefix
  \url{https://books.google.de/books?id=6benoAEACAAJ}.
\bibitem[{{International Association for the Properties of Water and
  Steam}(2014)}]{IAPWS2014}
\bibinfo{author}{{International Association for the Properties of Water and
  Steam}}, \bibinfo{title}{{IAPWS R1}-76(2014) {{R}}evised release on surface
  tension of ordinary water substance}, \bibinfo{year}{2014}. \URLprefix
  \url{http://www.iapws.org/relguide/Surf-H2O-2014.pdf}.
\bibitem[{{{Ravensberger Schmierstoffvertrieb GmbH}}(2022)}]{Ravenol2022}
\bibinfo{author}{{{Ravensberger Schmierstoffvertrieb GmbH}}},
  \bibinfo{title}{{{RAVENOL Getriebeöl CLP 220}}}, \bibinfo{year}{2022}.
  \URLprefix
  \url{https://www.ravenol.de/storage/app/media/product-pdf/Tds_1332109_en.pdf}.
\bibitem[{Ross(1950)}]{Ross1950}
\bibinfo{author}{S.~Ross}, \bibinfo{title}{Variation With Temperature of
  Surface Tension of Lubricating Oils}, \bibinfo{type}{Technical Report},
  United States National Advisory Committee for Aeronautics,
  \bibinfo{year}{1950}. \URLprefix
  \url{https://apps.dtic.mil/dtic/tr/fulltext/u2/a801538.pdf}.
\bibitem[{{3M}(2022)}]{Novec2022}
\bibinfo{author}{{3M}}, \bibinfo{title}{{$\text{3M}^\text{TM}\text{
  }\text{Novec}^\text{TM}$ 7500 Engineered Fluid}}, \bibinfo{year}{2022}.
  \URLprefix
  \url{https://multimedia.3m.com/mws/media/65496O/3m-novec-7500-engineered-fluid.pdf}.
\bibitem[{Brosseau et~al.(2014)Brosseau, Vrignon, and Baret}]{Brosseau2014}
\bibinfo{author}{Q.~Brosseau}, \bibinfo{author}{J.~Vrignon},
  \bibinfo{author}{J.-C. Baret},
\newblock \bibinfo{title}{Microfluidic dynamic interfacial tensiometry
  (\textmu{}{{DIT}})},
\newblock \bibinfo{journal}{Soft Matter} \bibinfo{volume}{10}
  (\bibinfo{year}{2014}) \bibinfo{pages}{3066--3076}. \URLprefix
  \url{http://dx.doi.org/10.1039/C3SM52543K}.
  \DOIprefix\doi{10.1039/C3SM52543K}.
\bibitem[{Denner and {van Wachem}(2015)}]{DENNER201524}
\bibinfo{author}{F.~Denner}, \bibinfo{author}{B.~G. {van Wachem}},
\newblock \bibinfo{title}{Numerical time-step restrictions as a result of
  capillary waves},
\newblock \bibinfo{journal}{Journal of Computational Physics}
  \bibinfo{volume}{285} (\bibinfo{year}{2015}) \bibinfo{pages}{24--40}.
  \URLprefix
  \url{https://www.sciencedirect.com/science/article/pii/S002199911500025X}.
  \DOIprefix\doi{https://doi.org/10.1016/j.jcp.2015.01.021}.
\bibitem[{Prosperetti(1981)}]{Prosperetti1981}
\bibinfo{author}{A.~Prosperetti},
\newblock \bibinfo{title}{Motion of two superposed viscous fluids},
\newblock \bibinfo{journal}{The Physics of Fluids} \bibinfo{volume}{24}
  (\bibinfo{year}{1981}) \bibinfo{pages}{1217--1223}. \URLprefix
  \url{https://aip.scitation.org/doi/abs/10.1063/1.863522}.
  \DOIprefix\doi{10.1063/1.863522}.
  \href{http://arxiv.org/abs/https://aip.scitation.org/doi/pdf/10.1063/1.863522}{{\tt
  arXiv:https://aip.scitation.org/doi/pdf/10.1063/1.863522}}.
\bibitem[{Lamb(1975)}]{lamb1975}
\bibinfo{author}{H.~Lamb}, \bibinfo{title}{Hydrodynamics},
  \bibinfo{edition}{6th} ed., \bibinfo{publisher}{Cambridge University Press},
  \bibinfo{year}{1975}.
\bibitem[{Shin and Juric(2002)}]{shin2002}
\bibinfo{author}{S.~Shin}, \bibinfo{author}{D.~Juric},
\newblock \bibinfo{title}{Modeling three-dimensional multiphase flow using a
  level contour reconstruction method for front tracking without connectivity},
\newblock \bibinfo{journal}{Journal of Computational Physics}
  \bibinfo{volume}{180} (\bibinfo{year}{2002}) \bibinfo{pages}{427--470}.

\end{thebibliography}
